\UseRawInputEncoding
\documentclass[aps,pre]{revtex4}

\usepackage{graphicx,amssymb,amsmath}
\usepackage{color}
\usepackage{multirow}

\newcommand{\be}{\begin{equation}}
\newcommand{\ee}{\end{equation}}

\newcommand{\bea}{\begin{eqnarray}}
\newcommand{\eea}{\end{eqnarray}}

\newcommand{\bzero}{\mathbf{0}}
\newcommand{\br}{\mathbf{r}}

\def\eq#1{Eq.~(\ref{#1})}
\def\eqs#1#2{Eqs.~(\ref{#1},\ref{#2})}
\def\fig#1{Fig.~(\ref{#1})}

\usepackage{array}
\newcolumntype{C}[1]{>{\centering\let\newline\\\arraybackslash\hspace{0pt}}m{#1}}
\newcolumntype{L}[1]{>{\raggedright\let\newline\\\arraybackslash\hspace{0pt}}m{#1}}
\newcolumntype{R}[1]{>{\raggedleft\let\newline\\\arraybackslash\hspace{0pt}}m{#1}}

\begin{document}

\title{Competition between Born solvation, dielectric exclusion, and Coulomb attraction in spherical nanopores}

\author{Th\'eo Hennequin}
\author{Manoel Manghi}
\email{manghi@irsamc.ups-tlse.fr}
\affiliation{Laboratoire de Physique Th\'eorique, Universit\'e de Toulouse, CNRS, UPS, France}
\author{John Palmeri}
\affiliation{Laboratoire Charles Coulomb (L2C), Universit\'e de Montpellier, CNRS, France}

\date{\today}

\begin{abstract}
The recent measurement of a very low dielectric constant, $\epsilon$, of water confined in nanometric slit pores leads us to reconsider the physical basis of ion partitioning into nanopores. For confined ions in chemical equilibrium with a bulk of dielectric constant $\epsilon_b>\epsilon$, three physical mechanisms, at the origin of ion exclusion in nanopores, are expected to be modified due to this dielectric mismatch: dielectric exclusion at the water-pore interface (with membrane dielectric constant, $\epsilon_m<\epsilon$), the solvation energy related to the difference in Debye-H\"uckel screening parameters in the pore, $\kappa$,
and in the bulk $\kappa_b$, and the classical Born solvation self-energy  proportional to $\epsilon^{-1}-\epsilon_b^{-1}$. Our goal is to clarify the interplay between these three mechanisms and investigate the role played by the Born contribution in ionic liquid-vapor (LV) phase separation in confined geometries. We first compute analytically the potential of mean force (PMF) of an ion of radius $R_i$ located at the center of a nanometric spherical pore of radius $R$.
Computing the variational grand potential for a solution of confined ions, we then deduce the partition coefficients of ions in the pore versus $R$ and the bulk electrolyte concentration $\rho_b$.
We show how the ionic LV transition, directly induced by the abrupt change of the dielectric contribution of the PMF with $\kappa$, is favored by the Born self-energy and explore the decrease of the concentration in the pore with $\epsilon$ both in the vapor and liquid states. Phase diagrams are established for various parameter values and we show that a signature of this phase transition can be detected by monitoring the total osmotic pressure as a function of $R$. For charged nanopores, these exclusion effects compete with the electrostatic attraction that imposes the entry of counterions into the pore to enforce electro-neutrality. This study will therefore help in deciphering the respective roles of the Born self-energy and dielectric mismatch in experiments and simulations of ionic transport through nanopores.
\end{abstract}

\maketitle

\section{Introduction}

Ion selectivity by synthetic membranes as well as in biological nanopores is known to be controlled by electrostatic interactions between the charged pore surface and mobile ions in solution~\cite{Schoch,revue_Levin,revue_John,Sym2007,Sym2009,Bocquet_charlaix,Kavokine}. A closer examination reveals that several mechanisms can be in competition in the transfer of an ion from a bulk reservoir to the interior of a pore: the electrostatic attraction of the counter-ions~\cite{Levitt,Levin}, as well as the exclusion induced by the possible dielectric mismatches between the confined electrolyte and membrane matrix, on the one hand, and between the confined and bulk electrolytes, on the other, the latter being at the basis of the so-called Born contribution.

The same mechanisms studied here enter into the partitioning of non-aqueous electrolytes (ionic liquids) into nanopores~\cite{Wu,Shrivastav,Vatamanu,Kondrat}, a topic of great current scientific and industrial interest, although the physical system parameters will take on different values compared to aqueous electrolytes. The theoretical approach we develop is general enough to be employed for both types of systems. A better theoretical understanding of both aqueous and non-aqueous electrolytes is necessary not only for reaching a deeper fundamental understanding of these rich physical systems, but also for optimizing these systems for a variety of societally important industrial applications.

Characterization of the impact of dielectric permittivity on ion concentration inside pores would also be useful for reaching a better understanding of technologies that are currently being investigated for the production of clean energy, for example those using  nanoporous carbon electrodes for blue energy~\cite{Chmiola,Simoncelli}, capacitive mixing~\cite{Brogioli} and capacitive deionization~\cite{Suss}. Indeed the capacitance of pores in materials such as carbide-derived carbon electrodes are directly dependent on the permittivity of the electrode and the solvent.

The dielectric mismatch between the confined electrolyte and the membrane matrix generates repulsive surface polarization charge in the case where the confined electrolyte has a higher dielectric constant (an effect simply interpreted in terms of image charge forces for planar
geometries)~\cite{Parsegian,Dresner1974,Yaroshchuk2000,Boda2007,PRE2010,PRL2010,JCP2011,Freger2018}. Such an effect can only be taken into account by going beyond mean-field approaches since the dielectric exclusion effect is embedded in the potential of mean force (PMF) via the excess chemical potential. Different electrolyte concentrations in the bulk and in the pore can also modify the solvation energy and therefore PMF, more precisely the value of the Debye screening  parameter $\kappa$ (inverse screening length) if correlations are taken at the Debye-H\"uckel (DH) level~\cite{DH}.

Similarly, the Born contribution to the solvation energy requires going beyond mean-field theory because it is also a dielectric effect that enters in the PMF via the excess chemical potential. Indeed, it has been known since the pioneering work of Stern~\cite{Stern} that to model experimental capacitance measurements reliably~\cite{Hunter,Lyklema}
it is necessary to consider that close to surfaces water has a reduced dielectric constant on a nanometric width. The dielectric constant of confined water has mainly been studied using molecular dynamics simulations~\cite{Ballenegger,Bonthuis2012}. These simulations indeed confirm that the out-of-plane water dielectric constant $\epsilon_\perp$ is lowered, close to surfaces, and that this decrease is associated with the local ordering of water molecules and the contribution of the multipoles (essentially quadrupoles and octupoles). This alignment induced by the presence of a surface induces a lower response of the water molecules to an applied electric field, and therefore a lower value of $\epsilon_\perp$ than in the bulk.
Several attempts have been made to model these effects using an extended Poisson-Boltzmann approach (which includes a spatially varying dielectric term $\nabla\epsilon_\perp^{-1}$) to describe the double-layer close to surfaces in water, and using an ad hoc chemical potential $\mu$ that mimics the wall repulsion, but neglects the electrostatic correlations between ions~\cite{Bonthuis2012,Loche}.
Attempts have also been made to assess the importance of dielectric effects in nanofiltration modeling without ~\cite{Dresner1974,Yaroshchuk2000} and with the Born contribution~\cite{Sym2007,Sym2009}.

In 2018, Fumagalli \textit{et al.}~\cite{Fumagalli} were able to measure, using local capacitance measurements, the out-of plane dielectric constant $\epsilon_\perp$ of water in  slabs of heights $h$ down to 1~nm bounded by hBN (of dielectric constant 3.5) with a very well controlled geometry. They measured a limiting value of $\epsilon_\perp\simeq 2.1$ for $h<2$~nm, and recovered the bulk value for $h>100$~nm. Their data were well fitted by a simple model of three capacitors in series made of a bulk layer (with a bulk value $\epsilon_\perp\simeq80$) sandwiched between two thin layers of width $0.74$~nm close to the surface. This low value of $\epsilon_\perp$ may arise simply from an averaging that includes the thin layers of vacuum close to hydrophobic surfaces (of width equal to twice the van der Waals radius of wall atoms)
with $\epsilon=1$, and a bulk value quickly reached in between (even for nanometric water slabs), as has been suggested by all-atom simulations~\cite{Zhang}.

A question which arises from these experiments is whether a low confined water dielectric constant plays a major role in the solvation energy barrier for ions entering in a pore of nanometric thickness. Indeed, when an ion of charge $q=ze$, where $z$ is the valence and $e$ the electron charge, is transferred from a high dielectric bulk region (with dielectric constant $\epsilon_b$) to a lower one (with dielectric constant $\epsilon$), Born~\cite{Born} showed in 1920 that a solvation energetic barrier, $W_{\rm Born}$ (in units of $k_BT$ with $T$ the temperature and $k_B$ the Boltzmann constant), exists and is given by
\be
W_{\rm Born} =\frac{q^2 \beta}{8\pi\epsilon_0 R_i} \left(\frac{1}{\epsilon}-\frac{1}{\epsilon_b}\right) =\frac{z^2\ell_B}{2 R_i} \left(\eta-1\right)
\label{born}
\ee
where $\eta = \epsilon_b/\epsilon$ measures the dielectric mismatch between bulk and confined water and $\ell_B=e^2/(4\pi\epsilon_0\epsilon_b k_BT)\simeq 0.7$~nm is the Bjerrum length in bulk water ($\epsilon_b=78$) at room temperature [$\beta=(k_B T)^{-1}$]. The effective ionic radius $R_i$ is approximately equal to the hard-core (or Pauling) ionic radius plus the water molecule ``radius''. $W_{\rm Born}$ is on the order of 1 for Na$^+$ ($R_i\simeq$~nm) if $\epsilon\simeq50$, but increases quickly with the square of the ion valence $z$. The precise value for $\epsilon$ that should be used in \eq{born} is not known. In the following, we will therefore study the whole range of $\epsilon$, from 2 to $\epsilon_b$. We recall that the Born solvation energy plays a dominant role in determining ion hydration from air (or vacuum) for which $\epsilon=1$ and therefore in this case $W_{\rm Born} \sim 100 - 200$, depending on ionic radius (see~\cite{JCP_Lory,Marcus,Rashin,Babu,Schmid,Marat} and references therein).

Following the pioneering work of Parsegian~\cite{Parsegian}, the role of this Born self-energy has been studied for the crossing of a flat dielectric interface~\cite{Boda} in the context of the KcsA K$^+$ channel~\cite{Liu} by correcting the Poisson-Nernst-Planck equations, or to study the binding selectivity of the L-type calcium channel~\cite{Boda2}. In this last paper, the authors use Grand Canonical Monte Carlo simulations to compare the energy barrier in the presence of low $\epsilon$ for monovalent Na$^+$ and divalent Ca$^{2+}$ ions with the case where $\epsilon=\epsilon_b$. Interestingly, they found no major difference, attributing this to a compensation between the increase of the Born self energy and the decrease of the electrostatic self energy when $\epsilon$ decreases. This numerical study, however, focused on a specific biological channel. Kiyohara et al.~\cite{Kiyohara} studied numerically the effect of the dielectric constant on electrolytes in porous electrodes with applied voltage. They essentially showed that decreasing the dielectric constant increases the electrostatic interaction between ions and explained the observed behavior in terms of the balance between the electrostatic interaction and the volume exclusion interaction. In these two articles based on numerical methods, despite their interest, a detailed physical understanding of these different contributions to the PMF is missing.

In this article, we develop a unified  analytical theory that considers on an equal footing the Born solvation, the dielectric exclusion, and the ionic correlation free energies. We then evaluate the role played by each in the ionic liquid-vapor phase transition previously proposed in the absence of the Born contribution~\cite{PRL2010,JCP2011,JCP2016}. We do not explicitly take into account steric exclusion because in the present context, the simplest way to do so  would involve reinterpreting the pore radius introduced here as an effective one equal to the nominal pore radius minus the ionic one, leading to a multiplicative factor in the ionic partition coefficients.  We recall that this previous work predicts that the confinement of a common mineral salt to a nanopore can lead to an ionic liquid-vapor phase transition at room temperature, in contrast to what occurs in the bulk (for which the theoretical transition temperature is predicted to be far below freezing and therefore unphysical).
In contrast to most studies of non-aqueous electrolytes (ionic liquids), where the shift in vapor-liquid coexistence induced by confinement is studied in the density-temperature plane~\cite{Wu}, we work at room temperature and investigate how ionic vapor-liquid coexistence is modified by the nanopore radius and bulk reservoir concentration.

To make the calculation tractable and therefore more clearly elucidate the basic physics at play, we choose to focus on a simplified geometry, namely a spherical nanopore~\cite{Dresner1974}, and go beyond the mean-field Poisson-Boltzmann approach by using a field theoretic variational theory~\cite{Netz2003,PRE2010} already developed for point-like ions in various  geometries such as slits~\cite{PRE2010,Lau}, spherical~\cite{Curtis2005}, and cylindrical~\cite{PRL2010,JCP2011} nanopores (finite sized ions including hard core steric effects were also studied approximatively in~\cite{JCP2016}, although in the absence of the Born contribution).
Taking into account fluctuations around the mean-field Poisson-Boltzmann solution amounts to including the contribution of the ionic self-energy in the theory~\cite{Wang,Liu2,Xu,Xua,Frusawa,Su} (including some attempts to account for the Born contribution).
In Ref.~\cite{Curtis2005} the field theoretic variational theory method was used to study electrolytes \textit{excluded} from, but not \textit{confined} to spherical regions (the focus here).  A complementary splitting-field method was also used in conjunction with Monte Carlo simulations to study electrolytes \textit{inside} spherical pores~\cite{Lue2015}, although the focus was different from our own and the question of the ionic liquid-vapor phase transition was not addressed.
It is crucial to underline that the field theoretic variational method developed here, unlike earlier methods~\cite{Yaroshchuk2000}, gives access to an approximate free energy (variational grand potential) functional that can be used to establish ionic liquid-vapor phase diagrams.

For simplicity and as a first approach to the complicated problem studied, we assume in this work that the solvent, the membrane, and the pore surface can be treated as continuous and homogeneous media characterized by continuous dielectric constants and  pore surface charge density.

In Section~\ref{single}, the full excess chemical potential is computed for a single ion located at the center of the pore, and its three contributions, namely dielectric, ionic  solvation and Born ones are compared. Then the variational theory is developed in Section~\ref{variational} to properly take into account ion-ion correlations at the Debye-H\"uckel level, in the midpoint approximation, i.e. the excess chemical potential for an ion located anywhere in the nanopore is taken to be equal to the one computed in Section~\ref{single} for an ion located at the pore center. Section~\ref{partition} is devoted to the computation of the partition coefficient of ions in the pore. A phase transition, mainly induced by the dielectric jump and already obtained for point-like ions~\cite{PRL2010,JCP2011}
and also for finite-sized ones within a restricted approximation~\cite{JCP2016}, is found between a state where no ions enter the pore (vapor phase) and a state where the pore concentration is more or less equal to the bulk one (liquid phase). The role of a low confined water dielectric constant on this transition is thoroughly studied. A discussion of the results and a conclusion are given in Section~\ref{discussion}.

\section{Solvation energy of a single ion in a spherical nanopore}
\label{single}

We consider a spherical nanopore of radius $R$ filled with a solvent with dielectric constant $\epsilon$. We use the Born model for ions which assumes that solvent is excluded from a spherical region of radius $R_i$ around a point charge $q= z e$ (restricted primitive model)~\cite{DH}. We call this Born radius $R_i$ the effective ionic radius, although it is approximately equal to the hard-core (or Pauling) ionic radius plus the water ``radius'' taken to be $R_{\rm water}\approx 0.142$~nm~\cite{Marcus}. One also could define this effective radius as the value for which the pair correlation function between the ion and the first water molecule reaches 1~\cite{JCP_Lory}. For Na$^+$ and Cl$^-$, the latter definition leads to $R_i\approx 0.2$~nm, which yields the correct values for the Born solvation energies \eq{born} when compared to molecular dynamics simulations. Although the former definition leads to larger values for the effective radii of these two ions, and therefore less accurate predictions for the Born solvation energies, qualitatively speaking we can conclude that the appropriate range is between 0.2 and 0.3~nm.

\begin{figure}[t]
\begin{center}
\includegraphics[height=6cm]{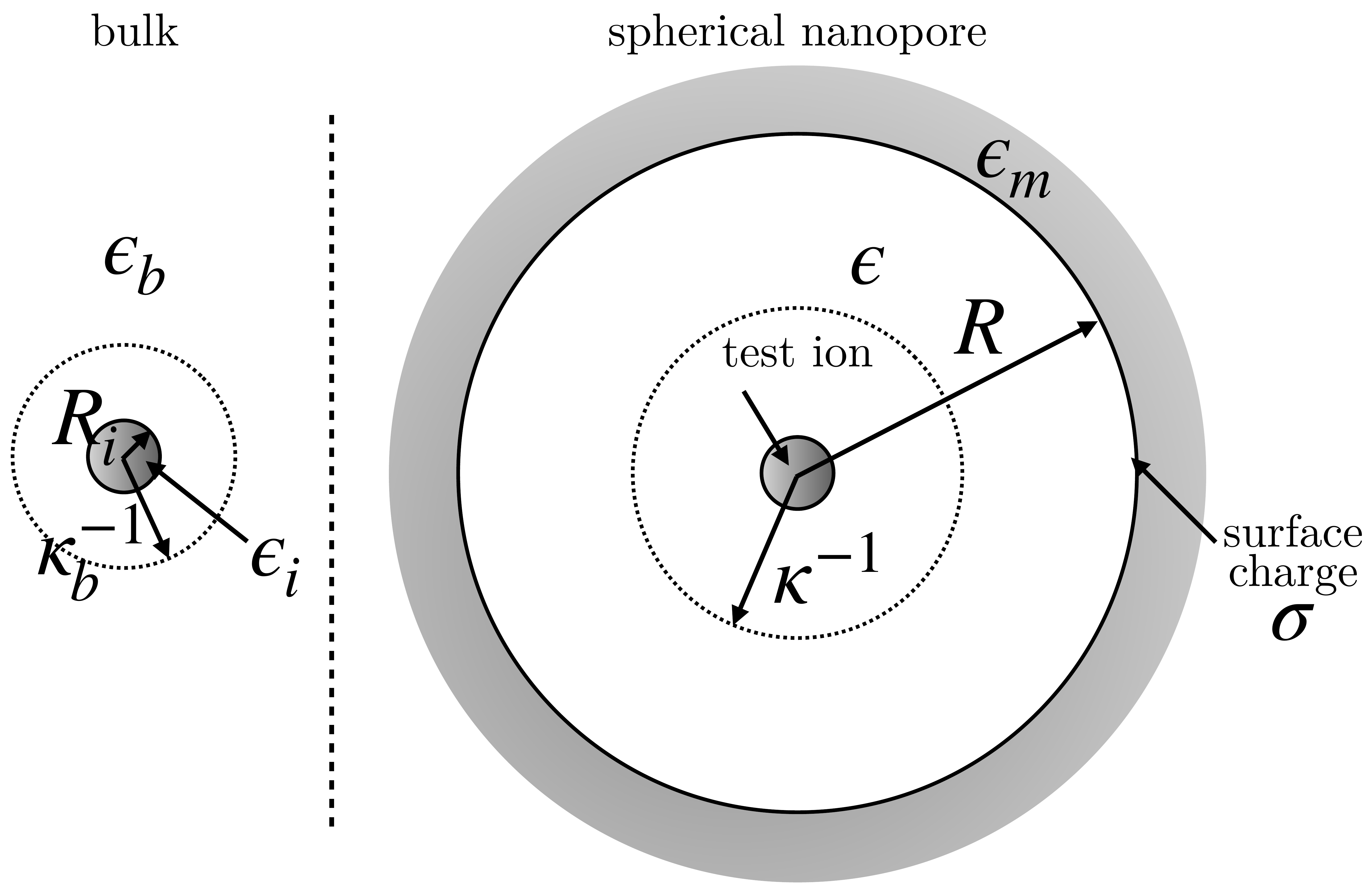}
\caption{Sketch of the geometry: a test ion of effective radius $R_i$ with internal dielectric constant $\epsilon_i$ is transferred from the bulk (dielectric constant $\epsilon_b=78$, Debye-H\"uckel constant $\kappa_b$) to the center of a spherical nanopore of radius $R$ embedded in a membrane of dielectric constant $\epsilon_m=2$. The confined water has a lower dielectric constant than in the bulk, $\epsilon\leq\epsilon_b$, and due to dielectric exclusion, the Debye-H\"uckel screening parameter is smaller than in the bulk, i.e., $\kappa\leq\kappa_b$. The internal nanopore surface possibly carries a surface charge density $\sigma$.}
\label{f0}
\end{center}
\end{figure}

This spherical region corresponds to the excluded volume of the ion and has a dielectric constant $\epsilon_i$ (Fig.~\ref{f0}). The membrane in which the nanopore is formed has dielectric constant $\epsilon_m$. This model can also be relevant for  the macropore-nanopore interfaces present in electrodes, where the macropores play the role of an effective bulk medium.

For a general \textit{linear} dielectric medium (assumed to be the case here) the total (normalized) electrostatic energy of a charge distribution is formally given by~\cite{Jackson}
\be
 W =\frac{\beta}{2}  \int d^3x \, \rho_c(\mathbf{x}) \Phi(\mathbf{x}),
\ee
where $\Phi(\mathbf{x})$ is the total potential (due to all charges) and $\rho_c(\mathbf{x})$ the total charge density (which may include point charges, as well as continuous charge distributions). For point charges $W$ contains infinite bare \textit{self-energy} contributions that must be subtracted off to obtain physically meaningful results. For the case of a single finite size ion located at the origin of spherical pore embedded in a general spherically symmetric dielectric medium, possibly with a continuous spherically symmetric charge distribution, this subtraction leads to a \textit{finite relative ionic (normalized) self-energy}~\cite{Born,Hunter,JCP_Lory}
\be
 W_{\rm p}=\frac{\beta q}2 \lim_{r\rightarrow 0}[\Phi_<(r)-\Phi_i(r)]
\ee
where
\be
\Phi_{i}(r) = \frac{q}{4 \pi \epsilon_0\epsilon_i r}
\label{bipot}
\ee
is the bare ionic Coulomb potential in a hypothetical uniform dielectric medium with
$\epsilon = \epsilon_m =\epsilon_i$ and $\Phi_{<}(r)$ is the electrostatic potential within the ion, i.e., for $r < R_i$.
It is found by solving the Poisson equation
\be
\frac{1}{r^2}\partial_r(r^2\partial_r\Phi)= - \frac{q\delta(\br)}{\epsilon_0 \epsilon_i}
\label{Peq}
\ee
and the usual boundary conditions at interface between media 1 and 2 (with surface charge density $\sigma$): $\Phi_1=\Phi_2$ and $\epsilon_1 \partial_r \Phi_1=\epsilon_2\partial_r  \Phi_2-\frac{\sigma}{\epsilon_0}$ where $\sigma$ is the surface charge density on the pore surface. The solution to this equation for $r < R_i$ can be written as
\be
\Phi_{<}(r) = \Phi_{i}(r) + \Delta \Phi,
\ee
where $\Delta \Phi$ is independent of $r$ and translates the influence of the surrounding medium on the central ionic charge.

The relative ionic \textit{self-energy} can therefore be simplified to
\be
 W_{\rm p}  =  \frac{\beta q}2 \Delta \Phi.
\ee
In the absence of pore surface charge the solution to the Poisson equation gives
\be
\Delta \Phi = \frac{q}{4 \pi \epsilon_0} \left[ \frac{1}{R_i}\left(\frac{1}{\epsilon}-\frac{1}{\epsilon_i}\right)+ \frac{1}{R}\left(\frac{1}{\epsilon_m}-\frac{1}{\epsilon}\right)\right]
\ee
and therefore in this case
\be
 W_{\rm p}
  =  \frac{\beta q^2}{8 \pi \epsilon_0} \left[ \frac{1}{R_i}\left(\frac{1}{\epsilon}-\frac{1}{\epsilon_i}\right)+ \frac{1}{R}\left(\frac{1}{\epsilon_m}-\frac{1}{\epsilon}\right)\right].
\label{Born SE}
\ee
One notices the very similar expressions for both of these two terms, which are both associated with dielectrics jumps. The first term is the usual Born self-energy in an unconfined solvent of dielectric constant $\epsilon$ (corresponding to the limit $R\to\infty$), originating in the dielectric mismatch between the ion ($\epsilon_i$) and solvent ($\epsilon$). The second term is the dielectric self-energy, originating in the dielectric mismatch between the confined solvent ($\epsilon$) and membrane itself ($\epsilon_m$).
In the special case of an ion in the bulk ($R\to\infty$ and $\epsilon\to\epsilon_b$), the relative ionic (normalized) self-energy simplifies to
\be
 W_{\rm b}=\frac{\beta q^2}{8 \pi \epsilon_0}
 \left[ \frac{1}{R_i}\left( \frac{1}{\epsilon_i} - \frac{1}{\epsilon_b}\right)
 \right]
\label{Born SEb}
\ee
and the difference in relative (normalized) self-energy between an ion in a pore and in the bulk,
\bea
\Delta W_{\rm p}
& = & W_{\rm p} - W_{\rm b}, \nonumber\\
& = & \frac{\beta q^2}{8 \pi \epsilon_0} \left[ \frac{1}{R_i}\left(\frac{1}{\epsilon}-\frac{1}{\epsilon_b}\right)+ \frac{1}{R}\left(\frac{1}{\epsilon_m}-\frac{1}{\epsilon}\right)\right],
\label{dWp}
\eea
is independent of $\epsilon_i$ and controls ion partitioning into the pore from the reservoir. The first term is the usual Born solvation self-energy, $W_{\rm Born}$ [\eq{born}], and the second is the usual dielectric one. When $\epsilon_m < \epsilon$ and $\epsilon < \epsilon_b$, both mechanisms disfavor ion partitioning into the pore from the bulk reservoir.
In the case where the pore bares a surface charge density $\sigma$, one has the following additional contribution
\be
\Delta W_{\rm p,\sigma}=\frac{\beta qR\sigma}{2\epsilon_0\epsilon_m},
\label{sigma_contr}
\ee
which is simply the mutual electrostatic energy of the ion and the charged pore.

Generalizing the Debye-H\"uckel (DH) approach for bulk electrolytes, we now consider the case of a finite size test ion located at the center of a spherical pore
filled with a (symmetric) electrolyte of concentration $\rho$.
The potential in this case is found by solving in the pore the DH equation, instead of the Poisson one as done previously:
\be
\frac{1}{r^2}\partial_r(r^2\partial_r\Phi) - \kappa^2 \Phi = 0,  \qquad  (R_i < r <R)
\label{DHeq}
\ee
with $\kappa=\sqrt{8\pi\eta\ell_B z^2 \rho}$ the pore DH screening parameter and the usual boundary conditions at interfaces between two media (see above). The Poisson equation, \eq{Peq}, still holds elsewhere in the system.

Before studying the general case, we recall the bulk result by taking
$\rho\to\rho_b$ (bulk electrolyte concentration), $R\to\infty$ and $\epsilon\to\epsilon_b$, in which case $\kappa\to\kappa_{\rm b}=\sqrt{8\pi\ell_B z^2 \rho_b}$, which leads to
\be
\Delta \Phi_{\rm b} = \frac{q}{4 \pi \epsilon_0 R_i}
\left[\left(\frac{1}{\epsilon_b}-\frac{1}{\epsilon_i}\right) - \frac{\kappa R_i}{1+ \kappa R_i}\right] =  \Delta \Phi_{\rm b}^{\rm B} + \Delta \Phi_{\rm b}^{\rm DH}
\ee
The (normalized) relative bulk self-energy in the presence of an electrolyte filled pore is then given by $W_{\rm b}=\frac{\beta q}{2}\Delta \Phi_{\rm b}$.
Up until now we considered the central ion as an inserted test ion. We now follow DH and consider the central ion to be part of the electrolyte itself and make the major assumption that the electrostatic part of the electrolyte \textit{excess} ionic chemical (normalized by $k_B T$) is given by $\mu^{\rm el}_{\rm b} = W_{\rm b}$:
\be
 \mu^{\rm el}_{\rm b}=\frac{z^2 \ell_B}{2 R_i}\left[\left(1-\frac{\epsilon_b}{\epsilon_i}\right) -
\frac{\kappa R_i}{1+ \kappa R_i}\right]
= \mu_{\rm b}^{\rm B} + \mu_{\rm b}^{\rm DH}.
\label{mubulk}
\ee
The first term on the RHS is the Born self-energy and the second term is the classical DH excess electrostatic chemical potential~\cite{McQuarrie}. This results generalizes the DH calculation to the more general case where $\epsilon_i\neq\epsilon_b$.

The above identification between $\mu^{\rm el} = W$  is completely equivalent to the DH \textit{charging method} as usually presented~\cite{DH}.
This method consists in computing the normalized volumetric Helmholtz electrostatic free energy,
$f^{\rm el}_{\rm b} = \beta F^{\rm el}_{\rm b}/V$ using the charging rule,~\cite{McQuarrie}
\be
f^{\rm el}_{\rm b}  = 2 \rho_{\rm b} \beta q \int_0^1 d\lambda \, \Delta\Phi_{\rm b}(\lambda q),
\ee
which using \eq{mubulk} is rewritten as
\bea
f^{\rm el}_{\rm b}& = & 2 \rho_{\rm b}  \mu_{\rm b}^{\rm B}+ 2 \rho_{\rm b} \beta q \int_0^1 d\lambda \, \lambda \,
\Delta\Phi_{\rm b}^{\rm DH}(\lambda \kappa_{\rm b}) \\
& = & 2 \rho_{\rm b}  \mu_{\rm b}^{\rm B}+ f^{\rm DH}_{\rm b} (\kappa_{\rm b}) ,\label{fel}
\eea
where we have used  that $\Phi_{\rm b}^{\rm B}$ depends explicitly on $q$, but is independent of $\kappa_{\rm b}$, and  that $\Phi_{\rm b}^{\rm DH}$ depends on $q$ explicitly and implicitly via $\kappa_{\rm b}(q)$ (which varies linearly with $q$).

The electrostatic DH contribution is rewritten as a function of $\kappa_{\rm b}$ using $2\rho_{\rm b} \beta q^2 = \epsilon_0\epsilon_{\rm b} \kappa_{\rm b}^2$:
\be
f^{\rm DH}_{\rm b}(\kappa_{\rm b})
= \frac{\epsilon_0\epsilon_{\rm b}}{2q} \kappa_{\rm b}^2 \int_0^1 d\xi \,  \Delta \Phi_{\rm b}^{\rm DH}(\kappa_{\rm b} \sqrt{\xi})
=\frac{\epsilon_0\epsilon_{\rm b}}{q}  \int_0^{\kappa_{\rm b}} \kappa  \,  \Delta \Phi_{\rm b}^{\rm DH}(\kappa) d\kappa
= -\frac{\kappa_{\rm b}^3}{12\pi} \tau(\kappa_{\rm b} R_i)
\ee
with
\be
\tau(x) = \frac{3}{x^3}
\left[
\ln (1+x) - x + \frac{x^2}{2}
\right].
\ee
We note that $f^{\rm DH}_{\rm b}(\kappa_{\rm b}) $ can be written directly in terms of the excess chemical potential:
\be
f^{\rm DH}_{\rm b}(\kappa_{\rm b})
=\frac{2\epsilon_0\epsilon_{\rm b}}{\beta q^2}  \int_0^{\kappa_{\rm b}} \kappa  \,  \mu_{\rm b}^{\rm el}(\kappa) d\kappa
=\frac{1}{2 \pi \ell_B  z^2}  \int_0^{\kappa_{\rm b}} \kappa  \,   \mu_{\rm b}^{\rm el}(\kappa) d\kappa.
\ee

The above result \eq{fel} for $f^{\rm el}_{\rm b}$ leads directly for a symmetric electrolyte to a consistent result for the chemical potential of the ionic species (cation or anion) $\mu^{\rm el}_{{\rm b},\pm}$:
\bea
\mu^{\rm el}_{{\rm b},\pm}
& = & \left( \frac{\partial f_{\rm b}^{\rm el}}{ \partial \rho_{{\rm b},\pm} }\right)_{V, T} = \left( \frac{\partial f_{\rm b}^{\rm el}}{\partial \rho_{{\rm b},\pm} }\right)_{\kappa_{\rm b}} +
\left( \frac{\partial f_{\rm b}^{\rm el}}{ \partial \kappa_{\rm b} }\right)_{\rho_{{\rm b,}s}}
\left( \frac{\partial \kappa_{\rm b} }{ \partial \rho_{{\rm b},\pm}  }\right)\\
&=&   \mu_{\rm b}^{\rm B} + \frac{\epsilon_0\epsilon_{\rm b}}{2q} \kappa_{\rm b} \Delta \Phi_{\rm b}^{\rm DH}(\kappa_{\rm b}) \frac{\kappa_{\rm b}}{4 \rho_{\rm b} } =  \mu^{\rm el}_{\rm b}
\eea
where we have used that in the present case $\kappa_{\rm b}$ can be written explicitly, using
$\rho_{\rm b} = (\rho_{\rm b+} + \rho_{\rm b-})/2$,
in terms of the ionic concentrations, $\rho_{\rm b+}$ and $\rho_{\rm b-}$, as $
\kappa_{\rm b} = \sqrt{4\pi\ell_B z^2 (\rho_{\rm b+} + \rho_{\rm b-})}$,
and therefore $\frac{ \partial \kappa_{\rm b} }{ \partial \rho_{{\rm b,}s} } =
\frac{\kappa_{\rm b}}{4 \rho_{\rm b} }$.

In general the full ionic chemical potential is obtained by adding the ideal gas entropic contribution to the electrostatic one:
\be
\mu_\pm  =\ln \rho_\pm + \mu^{\rm el}_\pm.
\ee

The normalized grand potential, $\omega = \beta\Omega/V = -\beta p$, where $p$ is the pressure, can be obtained directly from  thermodynamics, leading in the present (symmetric electrolyte) case to
\be
\omega = -\beta p = f - 2 \rho \mu,
\label{thomega}
\ee
since the total ionic concentration is $2 \rho$.
This relation leads to an explicit \textit{charging} form for the excess (normalized) electrostatic grand potential in the bulk (proportional to the pressure) :
\bea
\omega^{\rm el}_{\rm b}
& = & -\beta p^{\rm el}_{\rm b} = f^{\rm el}_{\rm b}
- 2 \rho_{\rm b}  \mu^{\rm el}_{\rm b}, \nonumber\\
& = & \rho_{\rm b}  \beta q
\left[
\int_0^1 d\xi \,  \Delta\Phi_{\rm b}(\kappa_{\rm b} \sqrt{\xi}) - \Delta\Phi_{\rm b}(\kappa_{\rm b})
\right],\\
& = & \frac{\epsilon_0\epsilon_{\rm b}}{2q}  \kappa_{\rm b}^2
\left[
\int_0^1 d\xi \,  \Delta\Phi_{\rm b}^{\rm DH}(\kappa_{\rm b} \sqrt{\xi}) - \Delta\Phi_{\rm b}^{\rm DH}(\kappa_{\rm b})
\right],\\
& = & \frac{1}{2 \pi \ell_B  z^2}  \int_0^{\kappa_{\rm b}} \kappa  \,
\left[
\mu_{\rm b}^{\rm el}(\kappa) - \mu_{\rm b}^{\rm el}(\kappa_{\rm b})
\right] d\kappa. \label{omegacp}
\eea
since the Born contribution, which is independent of $\kappa$, does not contribute to the electrostatic pressure, or
\be
\beta p^{\rm el}_{\rm b} (\kappa_{\rm b}) = -\frac{\kappa_{\rm b}^3}{24\pi}
\left[
\frac{3}{1+\kappa_{\rm b} R_i}-2\tau(\kappa_{\rm b} R_i)
\right].
\ee

We assume now that the nanopore is filled by an electrolyte with in general a \textit{different} DH screening parameter,
\be
\kappa=\sqrt{8\pi\ell_B\eta z^2 \rho},
\ee
from the one in the bulk, where $\rho$ is the electrolyte concentration in the pore (see \fig{f0}).
Moreover, we limit ourselves for the moment to the case $\sigma=0$.
Generalizing the above bulk calculation to the case of an ion embedded in a spherical nanopore geometry yields a more complicated, but analytical, expression for the electrostatic potential $\Phi_{\rm p} (r) = \Phi_i(r) + \Delta\Phi_{\rm p}(\kappa)$ (see Appendix).  (The volumetric free energy $f_{\rm p}$ obtained from the charging method leads to an integral that cannot, however,  be carried out analytically.) Using $\mu^{\rm el}_{\rm p}=\frac{\beta q}{2}\Delta \Phi_{\rm p}(\kappa)$, one obtains an analytical expression for the excess electrostatic chemical potential in the pore
\be
\mu^{\rm el}_{\rm p} =\frac{z^2 \eta \ell_B}{2 R_i}
\left[
\frac{2 \kappa R_i}{
\left(
e^{2 \kappa (R-R_i)}(1+\kappa R_i)\frac{\kappa R-1+\epsilon_m/\epsilon}{\kappa R+1-\epsilon_m/\epsilon}+1-\kappa R_i
\right)(1+\kappa R_i)
}+ \left(1 - \frac{\epsilon}{\epsilon_i}\right) - \frac{\kappa R_i}{1+\kappa R_i}
\right]
\label{mupore}
\ee
Taking $R\to\infty$ kills the first term in the brackets, and we recover \eq{mubulk} by replacing $\epsilon$ with $\epsilon_b$ (or $\eta=1$). We assume that the DH charging method, outlined above for the bulk case, can be carried over \textit{mutatis mutandis} to the pore case to obtain
$f^{\rm el}_{\rm p}$ and therefore
$\omega^{\rm el}_{\rm p}$ from
$\mu^{\rm el}_{\rm p}$ and $\rho$:
\be
f^{\rm el}_{\rm p}
 =   \beta\rho q \int_0^1 d\xi \,  \Delta\Phi_{\rm p}(\kappa \sqrt{\xi})
\ee
and
\be
\omega^{\rm el}_{\rm p}
 =  \frac{  \epsilon_0\epsilon}{2q} \kappa^2
\left(
\int_0^1 d\xi \,  \Delta\Phi_{\rm p}(\kappa \sqrt{\xi}) - \Delta\Phi_{\rm p}(\kappa)
\right),
\ee
where we have used $2\rho\beta q^2 = \epsilon_0\epsilon \kappa^2$.
If we assume that the pore interior is in equilibrium with a bulk external reservoir, then the two electrochemical potentials are equal,
$\mu_{\rm p} = \mu_{\rm b}$,
where
\be
\mu_{\rm p}  = \ln \rho + \mu^{\rm el}_{\rm p}
\ee
and
\be
\mu_{\rm b}  = \ln \rho_{\rm b} + \mu^{\rm el}_{\rm b}.
\ee
Therefore the partition coefficient is
$k=\rho/\rho_{\rm b}=\exp(-\Delta W_{\rm p})$,
where
\be
\Delta W_{\rm p}(\kappa) = \mu^{\rm el}_{\rm p}-\mu^{\rm el}_{\rm b},
\ee
is the difference in chemical potentials (or PMF). This PMF, which controls the transfer of an ion from the bulk to the center of the spherical nanopore, is given by
\bea
\Delta W_{\rm p}(\kappa) & = & W_{\rm conf}(\kappa,\epsilon_m/\epsilon)+W_{\rm DH}(\kappa)+W_{\rm Born}\\
&=& \frac{z^2\ell_B\eta\kappa}{e^{2 \kappa (R-R_i)}(1+\kappa R_i)^2\frac{\kappa R-1+\epsilon_m/\epsilon}{\kappa R+1-\epsilon_m/\epsilon}+1-(\kappa R_i)^2}  + \frac{z^2 \ell_B}{2}\left(\frac{\kappa_b}{1+ \kappa_b R_i}-\frac{\eta\kappa}{1+\kappa R_i}  \right)+ \frac{z^2 \ell_B}{2 R_i}(\eta-1)
\label{W}
\eea
In particular, on can check that when $\kappa$ and $\kappa_b\to0$,
$\Delta W_{\rm p}(\kappa)$ reduces to the correct result in the absence of electrolyte, \eq{dWp}, comprised of the usual  Born self-energy,
$W_{\rm Born}$, and dielectric solvation contributions.
In the case where the pore bares a surface charge density $\sigma$, one should add the following pore-charge electrostatic energy
\be
\Delta W_{\rm p,\sigma}(\kappa) = \frac{\beta q\sigma\kappa}{\epsilon_0\epsilon}\frac{e^{\kappa (R-R_i)}}{e^{2 \kappa (R-R_i)}(1+\kappa R_i)(\kappa R-1+\epsilon_m/\epsilon) +(\kappa R+1-\epsilon_m/\epsilon)(1-\kappa R_i)}
\ee
which leads to \eq{sigma_contr} in the limit $\kappa\to0$.

One can, somewhat artificially, separate the excess chemical potential given in \eq{W} into three contributions: (i)~the first term $W_{\rm conf}(\kappa,\epsilon_m/\epsilon)$ is associated with the confinement of ions in a nanopore and depends on the ratio of dielectric constants $\epsilon_m/\epsilon$ and the cavity radius $R$ (note that this term increases dramatically when $R_i$ increases but the divergence occurs for $R_i>R$, i.e. an unphysical case); (ii)~the second term $ W_{\rm DH}(\kappa)$ is the difference in solvation energies (related to the DH chemical potential for an ion of effective radius $R_i$) between a hypothetic bulk with DH constant $\kappa$ and the bulk with $\kappa_b$; and (iii)~the last term $W_{\rm Born}$ is independent of both $\kappa$ and $R$. It corresponds to the classical Born solvation energy of an ion which is transferred from the bulk to the pore with $\epsilon\neq\epsilon_b$ ($\eta\neq 1$). It dominates for small ion radii, and diverges for vanishing $R_i$ since, in this case, it corresponds to the difference of the Coulomb self-energy for point-like ions.
Hence for $\epsilon<\epsilon_b$, the Born solvation energy impedes the entrance of ions into the nanopore.
\begin{figure}[t]
\begin{center}
\includegraphics[height=5cm]{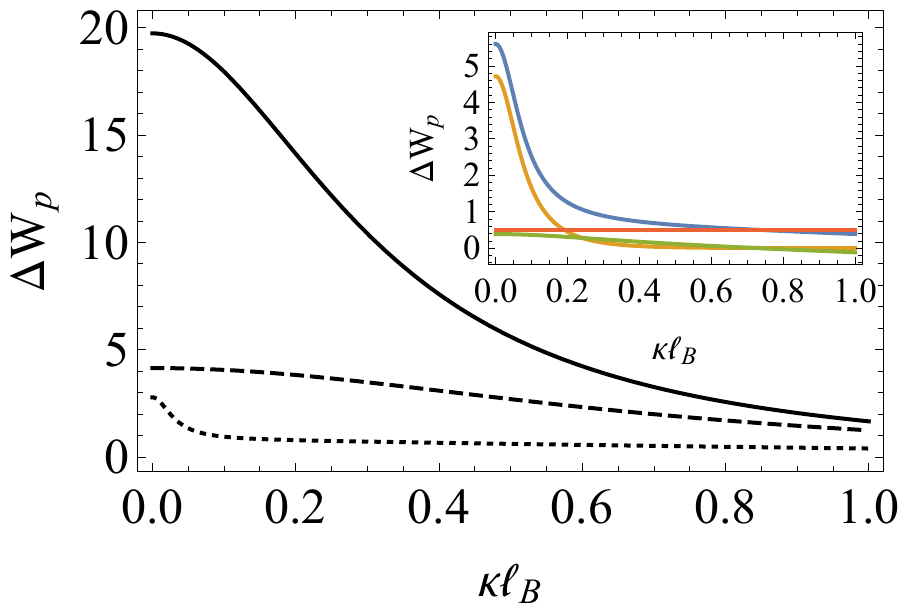} \hspace{2cm}\includegraphics[height=5cm]{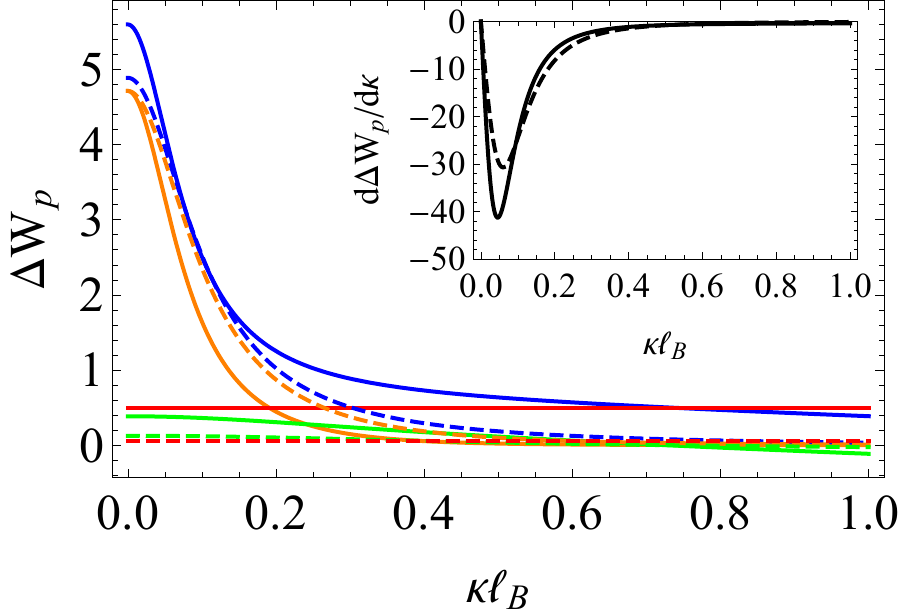}\\
(a)\hspace{9cm}(b)
\caption{Potential of mean force $\Delta W_{\rm p}$ (in units of $k_BT$) given in \eq{W} versus $\kappa\ell_B$ for $\kappa_b\ell_B=1$ (or $\rho_b \approx 0.19$~M), and $\epsilon=60$: (a)~$R_i=0.3\ell_B$ and $R=2\ell_B$, $\epsilon_m=2$ (solid line); $R=2\ell_B$, $\epsilon_m=10$ (dashed line), $R=10\ell_B$, $\epsilon_m=2$ (dotted lines).
Inset: various contributions to $\Delta W_{\rm p}$: dielectric $W_{\rm diel}$ (yellow), solvation $W_{\rm sol}$ (green), and Born $W_{\rm B}$ ones (red) for  $R=4\ell_B$, $\epsilon_m=2$ with the other parameters being the same.
(b) Influence of the ionic size  for $\kappa_b\ell_B=1$, $\epsilon=60$, $\epsilon_m=2$, $R=4\ell_B$, and $R_i=0.3\ell_B$ (solid lines); $R_i=3\ell_B$ (dashed lines). Inset: First derivative $W'(\kappa)$.}
\label{f1}
\end{center}
\end{figure}

It is physically illuminating to define the three following contributions to the PMF
\be
\Delta W_{\rm p}(\kappa) =  W_{\rm diel}(\kappa,\epsilon_m/\epsilon)+W_{\rm sol}(\kappa)+W_{\rm Born}
\ee
where $W_{\rm diel}(\kappa,\epsilon_m/\epsilon)=W_{\rm conf}(\kappa,\epsilon_m/\epsilon)-W_{\rm conf}(\kappa,1)$ is the dielectric jump one, $W_{\rm sol}=W_{\rm conf}(\kappa,1)+W_{\rm DH}(\kappa)$ the solvation one, and $W_{\rm Born}$, the  Born one.
They depend on 6 dimensionless parameters which are $\kappa_b\ell_B$,  $\kappa\ell_B$, $R/\ell_B$, $R_i/\ell_B$, $\epsilon_m/\epsilon_b$, and $\eta$. The variation of $\Delta W_{\rm p}(\kappa)$ and its different contributions with $\kappa\ell_B$ are shown in \fig{f1}(a) for
$\kappa_b\ell_B=1$ (or $\rho_b \approx 0.19$~M), $\epsilon=60$,
and $R_i=0.3\ell_B$ and different values of the other parameters. One clearly see that $\Delta W_{\rm p}$ decreases exponentially with $\kappa R$ and that its value at low $\kappa$ decreases with increasing $\epsilon_m$ or $R$, as dielectric exclusion diminishes. The inset in \fig{f1}(a) shows that the behavior of $\Delta W_{\rm p}$ is controlled by the dielectric exclusion, $W_{\rm diel}$, at small $\kappa$ and by the constant Born exclusion contribution, $W_{\rm Born}$, for $\kappa\simeq\kappa_b$. In any case, the solvation contribution, $W_{\rm sol}$, which accounts for the distortion of the ionic screening cloud in a confined geometry, is smaller than $k_BT$ and becomes slightly negative for $\kappa$ close to $\kappa_b$.

Finally \fig{f1}(b) shows the effect of the ionic size $R_i$ on $\Delta W_{\rm p}(\kappa)$ and its different contributions. For large ions ($R_i=3\ell_B$), the Born contribution decreases as expected and therefore $W\simeq W_{\rm diel}$ which remains strong for $\epsilon_m=2$. Interestingly, changing $R_i$ essentially shifts the PMF due to the change in the Born contribution, but does not modify so much its variations, and therefore its first derivative, as shown in the inset of \fig{f1}(b).

\section{Variational approach}
\label{variational}

In the preceding section, the charged particle at the pore center was first considered as a single ion in a dielectric cavity in the absence of electrolyte, possibly interacting with charged pore walls, leading to the solvation self-energy, $\Delta W_{\rm p}$, \eq{dWp}. The special case where  the pore surface charge exactly counterbalances that of the ion (global electro-neutrality) corresponds to so-called \textit{strong coupling limit}~\cite{Netz2001}. In this case to the lowest order approximation the ion-ion interactions are neglected and only the electrostatic interactions of ions with the pore walls are taken into account.

If the ion is embedded, however, in an electrolyte-filled pore, the solvation self-energy, which depends on electrolyte concentration, is computed using a screened electrostatic potential and leads to the PMF given by \eq{W}. In this case we have assumed that $\sigma=0$ and generalized the bulk DH calculation to an  electrolyte-filled pore where electro-neutrality is enforced by the ionic cloud surrounding the test charge.

In Section II, we have studied the self-energy of a single ion inside a nanopore with a given Debye constant $\kappa$. In this Section, we use this self-energy to determine self-consistently the optimized value of the Debye constant $\kappa_v$ using a variational approach and then use this value to obtain the optimized variational grand potential.

To study the many-body statistical physics of an electrolyte in a spherical nanopore by properly incorporating the dielectric, solvation and Born energies, we must go beyond the mean-field Poisson-Boltzmann approach and the lowest order strong coupling approach (developed in the previous section). To do so we use the Gaussian variational theoretical approach developed in~\cite{Netz2003,PRE2010,PRL2010,JCP2011,JCP2016}. The full Hamiltonian is
\be
\beta H= \int d\br \left[\frac{\varepsilon(\br)}{2 \beta e^2}(\vec{\nabla} \Phi(\br))^2 -i \rho_c (\br) \Phi(\br)-\sum_{j=\pm} \lambda_j e^{z_j^2 v_i(0)/2+i z_j \Phi(\br)} \right]
\ee
where $\Phi(\br)$ is a fluctuating field related to the fluctuating electric potential, $\rho_c (\br) = \sigma \delta(r - R)$ is the external (volumetric) charge density that accounts for the pore surface charge density, $\sigma$ (in units of $e$), and $v_i(\br)$ the bare ionic (dimensionless) Coulomb potential in a hypothetical uniform medium of dielectric constant $\epsilon_i$ (see \eq{bipot}),
\be
v_i(\br,\br') = \frac{\ell_B \epsilon_b}{\epsilon_i |\br-\br' |}
\ee
[$v_i(0)$ is therefore the infinite bare ionic self-energy]
and $\lambda_j$ the fugacity of ion of type $j$. This Hamiltonian is approximated by a Gaussian variational form,
\be
\beta H_0=\frac12\int_{\br,\br'}\left[\Phi(\br)-i \Phi_0(\br)\right] v_0^{-1}(\br,\br')\left[\Phi(\br')-i\Phi_0(\br')\right]
\ee
where $\Phi_0(\br) = -i \left\langle \Phi(\br) \right\rangle_0$ is the mean physical (dimensionless) variational electrostatic potential (for the variational Hamiltonian) induced by the fixed charges (e.g., the surface charge density of the walls) and $v_0(\br,\br')$ is the (dimensionless) electrostatic kernel governing the interaction between test ions located at $\br$ and $\br'$. It depends directly on the dielectric distribution, but only indirectly on the surface charge $\sigma$ via $\Phi_0(\br)$. Hence the two electrostatic interactions are partially decoupled via their respective sources: $\Phi_0(\br)$ has as source only the fixed electrostatic charges and $v_0(\br,\br')$ has as source only the point test ion charges.

For sake of simplicity, we assume that in a small nanopore, $\Phi_0$ can be approximated by a constant Donnan potential and is therefore determined by global electro-neutrality in the pore. We first consider only point-like ions (i.e., $\epsilon_i = \epsilon$) and assume that the corresponding $v_0$ is the Green function solution to a DH equation with a constant variational DH screening parameter $\kappa_v$,
\be
v_0(\br,\br';\kappa_v) = v_0^{\rm DH}(\br,\br';\kappa_v)+ \delta v_0(\br,\br';\kappa_v),
\label{v0v}
\ee
where $\delta v_0(\br,\br')$ is defined as the correction to an effective bulk variational (dimensionless) DH potential in a medium with dielectric constant $\epsilon$,
\be
v_0^{\rm DH}(\br,\br';\kappa_v) =
\eta \ell_B  \frac{e^{- \kappa_v  |\br'-\br| }}{|\br-\br'|}.
\ee
Clearly the correction $\delta v_0(\br,\br';\kappa_v)$  takes into account the presence of dielectric inhomogeneities and therefore depends on the pore geometry. It can be evaluated exactly for cylindric~\cite{JCP2011} and spherical~\cite{Curtis2005} pores (see Appendix). The dielectric mismatch parameter $\eta$ has been introduced to handle the case where $\epsilon\neq\epsilon_b$.

For point-like ions the variational parameters $\kappa_v$ and $\Phi_0$ can then be  obtained by minimizing the  dimensionless volumetric variational grand-potential $\omega_{\rm v}=\beta\Omega_{\rm v}/V$ where,
\be
\Omega_{\rm v} = \Omega_{0} + \left\langle H-H_0 \right\rangle_0,
\ee
and the expectation value is taken with $H_0$.
Evaluating $\Omega_{\rm v}$ using \eq{v0v} leads to~\cite{JCP2011}
\bea
\omega &=& -\sum_{j = \pm}  \left\langle \rho_j (\br) \right\rangle  + \frac{\kappa_v^3}{24 \pi}  \nonumber \\
    &&+ \frac{\kappa_v^2}{8 \pi \ell_B \eta} \int_0^1 d\xi \langle\delta v_0(\br,\br;\kappa_v \sqrt{ \xi } )-\delta v_0(\br,\br;\kappa_v)\rangle+\frac{3}{R}\sigma \Phi_0,
    \label{var_point}
\eea
where $V=\frac43 \pi R^3$ is the pore volume,
\be
\left\langle \rho_\pm (\br) \right\rangle =
\lambda_\pm
\left\langle
e^{-w_\pm (\br)}
\right\rangle
\ee
is the pore averaged concentration of cations ($+$) and anions ($-$) (the symbols $\langle\ldots\rangle$ stands for an average over the nanopore volume),
which can be identified using
$\rho_{\pm}=-\lambda_\pm \frac{\partial \omega}{\partial\lambda_\pm}$ and
\be
w_\pm(\br) = \frac{z^2}{2}
\left[
\delta v_0(\br,\br;\kappa_v) - \eta\kappa_v \ell_B
\right]
\pm z\Phi_0
\ee
with
\be
\lambda_\pm=\rho_{b} \exp({-z^2\kappa_b\ell_B/2})
\ee
the fugacity of ions of type $\pm$ (determined by the bulk reservoir with which the ions in the pore are in equilibrium). Since we have limited our approach to symmetric electrolytes with $z_+=z_-=z$, $\lambda_+=\lambda_-=\lambda$ (the generalization to asymmetric salts is straightforward but leads to more complicated equations).
The two first terms on the RHS of \eq{var_point} are equal to
$-\beta p$,
where $p =  p^{\rm id} + p^{\rm el}$ is an effective pressure, composed of the ideal osmotic pressure (first term) and the electrostatic osmotic DH pressure of point ions, $-\kappa_v^3/(24\pi)$ (second term), for a hypothetical \textit{bulk} with DH screening parameter equal to $\kappa_v$.
The first (negative ideal osmotic pressure) term with a PMF $w(\br)$ includes $\delta v_0(\br,\br)$ and $\Phi_0$, since the ionic concentrations are modified by the presence of the pore walls. The two last terms are surface terms, equal to $\frac{3}{R}\beta\gamma$ ($\frac{3}{R}=S/V$ for a sphere), where $\gamma$ is the surface tension, that include the dielectric ($\delta v_0$) and electrostatic ($\Phi_0$) contributions induced by the presence of the pore wall.
The second \textit{bulk-like} term and the third surface term in \eq{var_point}, which are computed using the variational DH Green function, $v_0 =
v_0^{\rm DH}+\delta v_0$, are identical to the ones obtained following the DH charging method~\cite{McQuarrie}:
\be
\frac{\kappa_v^2}{8 \pi \ell_B \eta} \int_0^1 d\xi
\langle
v_0(\br,\br;\kappa_v \sqrt{ \xi } )
-v_0(\br,\br;\kappa_v)
\rangle
\ee

In the bulk ($\delta v_0=\Phi_0=0$, $\epsilon=\epsilon_b$), it has been shown using this variational approach that the variational DH screening parameter $\kappa_v$ reduces to its usual bulk value for low enough electrolyte concentration~\cite{JCP2011}
\be
\kappa_b^2=4\pi\ell_B\sum_{j=\pm} z^2_j\lambda_j\exp\left(\frac{z_j^2}{2}\kappa_b \ell_B\right)=4\pi\ell_B\sum_{j=\pm}  z^2_j\rho_{j,b}
\label{kappab}
\ee

In the case where $R\to\infty$ and the  pore is uncharged ($\delta v_0=\Phi_0=0$, $\epsilon\neq\epsilon_b$), the minimization of \eq{var_point} leads to the following variational equation
\be
\kappa_v^2=4\pi\ell_B\eta \sum_{j=\pm}  z^2_j\lambda_j\exp\left(\frac{z_j^2}{2}\eta\kappa_v \ell_B \right)=4\pi\ell_B\eta\sum_{j=\pm}  z^2_j \rho_{j}
\label{kappav}
\ee
which is the usual DH screening parameter for a bulk electrolyte of dielectric constant $\epsilon$.

In calculating the variational grand potential both \textit{bulk-like} and \textit{surface} contributions to the ionic self-energy [arising from $\delta v_0(\br,\br)$] must be accounted for in \eq{var_point}. Hence to take into account the ion finite size in the electrostatic self-energy, and therefore the dielectric jumps both at the pore surface and the ion surface, we have to modify $\delta v_0(\br,\br)$. In the following we modify \eq{var_point} by (i)~taking into account the ion finite size in the DH electrostatic potential (and  free energy and chemical potential), and (ii)~using the mid-point approximation~\cite{Yaroshchuk2000,PRE2010}, i.e. computing the electrostatic self-energy only at the center of the pore, which therefore provides a lower bound for the dielectric exclusion, which is higher closer to the pore wall. We focus here on already subtle electrostatic effects and therefore do not consider the more complicated case of hard-core repulsion, which can also be included in this variational approach using a Carnahan-Starling formula for the pressure and plays a non-negligible role at high concentrations (roughly for $\rho_b>0.5$~mol/L)~\cite{JCP2016}.

Following points (i)-(ii) listed above, the difference in chemical potentials of point-like ions in a spherical nanopore
\be
\mu(\br)-\mu_b=\frac{z^2}{2}[\delta v_0(\br,\br,\kappa_v)+(\kappa_b-\eta\kappa_v) \ell_B ]
\ee
is therefore modified by taking $\br$ at the pore center in the average over pore volume (midpoint approximation) and replaced by the difference of chemical potentials, $\Delta W_{\rm p}(\kappa_v)$, given in \eq{W} to take into account the finite size of the ions. Hence $z^2\delta v_0/2$ is replaced by $ W_{\rm conf}(\kappa_v,\epsilon_m/\epsilon)$, and
$z^2\ell_B(\kappa_b-\eta\kappa_v)/2$ by $W_{\rm DH}(\kappa_v) + W_{\rm Born}$.

The second and third terms of \eq{var_point} are computed using the charging method presented in the previous section, using \eq{thomega}, which leads to:
\bea
\omega(\kappa_v,\Phi_0) &=& -2\rho_b e^{- \Delta W_{\rm p} (\kappa_v)}\cosh(z\Phi_0) + \frac1{8\pi R_i^3} \left[2 \kappa_v R_i-2 \ln(1+\kappa_v R_i)-\frac{\kappa^2_v R^2_i}{1+\kappa_v R_i}\right]\nonumber \\
    &&+ \frac{\kappa_v^2}{4 \pi z^2\ell_B\eta}\left[\int_0^1 W_{\rm conf}(\sqrt{\xi}\kappa_v)d\xi -W_{\rm conf}(\kappa_v)\right]+\frac{3}{R}\sigma \Phi_0
\label{omega}
\eea
As for point ions, the Born contribution enters in $\Delta W_{\rm p}$, but not the effective bulk-like pressure contribution.
It is useful to write $\omega(\kappa_v,\Phi_0)$ entirely in terms of the difference in excess chemical potentials, $\Delta W_{\rm p}$, using $\kappa$ as the charging parameter [cf. \eq{omegacp}]:
\be
\omega(\kappa_v,\Phi_0) = -2\rho_b e^{- \Delta W_{\rm p} (\kappa_v)}\cosh(z\Phi_0)
+ \frac{1}{2\pi z^2\ell_B\eta}
\int_0^{\kappa_v}    d\kappa \kappa
\left[
\Delta W_{\rm p}(\kappa) -\Delta W_{\rm p}(\kappa_v)
\right]
+\frac{3}{R}\sigma \Phi_0.
\label{omega1}
\ee
The second term on the RHS of \eq{omega1} constitutes the excess variational electrostatic grand potential, $\omega^{\rm el}(\kappa_v)$.

Minimizing \eq{omega1} with respect to the variational parameters $\Phi_0$ and $\kappa_v$ allows us to obtain the equilibrium expectations values. In the following, we focus on the ionic concentrations in the nanopore.
The minimization with respect to $\kappa_v$ allows us to obtain immediately, owing to the mid-point approximation (cf. Eq.~1 of \cite{JCP2011})
and the relation
\be
\frac{\partial\omega^{\rm el}}{\partial\kappa_v} =
-\frac{\kappa_v^2}{4\pi z^2\ell_B\eta}
\frac{\partial\Delta W_{\rm p} }{\partial\kappa_v},
\label{domega}
\ee
the following simple result:
\be
\kappa_v^2 = 8 \pi \ell_B \eta z^2\rho_b e^{- \Delta W_{\rm p}(\kappa_v)} \cosh(z \Phi_0)=4\pi\ell_B\eta z^2(\rho_++\rho_-),
\label{eq:kappaminim}
\ee
where
\be \rho_\pm=\rho_b e^{- \Delta W_{\rm p} \mp z\Phi_0}
\ee
are the concentrations in the pore and
\be
k_\pm \equiv \frac{\rho_\pm}{\rho_b}= e^{- \Delta W_{\rm p}(\kappa_v)\mp z\Phi_0}
\ee
the partition coefficients. The relation \eq{domega} is simply the variational analog of the usual thermodynamic identity
$(\partial\omega/\partial\rho) = - 2 \rho (\partial\mu/\partial\rho)$ for a two component system, here a symmetric electrolyte with a total ionic concentration of $2\rho$.

The minimization with respect to the Donnan potential $\Phi_0$ leads to the electroneutrality condition:
\be
\sigma = \frac23 zR \rho_b  e^{- \Delta W(\kappa_v)} \sinh(z \Phi_0)=\frac{zR}3(\rho_--\rho_+)
\label{electro}
\ee
Two equations similar to \eqs{eq:kappaminim}{electro}, but more general, were obtained for point-like ions in a cylindrical geometry~\cite{JCP2011} (without the mid-point approximation).

In particular, in the bulk limit ($R\to \infty$) for $R_i\neq0$, we find the usual DH Finite Size result for the chemical potential in a solution~\cite{McQuarrie}, which, for $\eta\neq1$, is generalized to include the Born contribution:
\be
\mu=\ln\rho - \frac{z^2 \eta\ell_B}{2R_i}
\left[
\frac{\kappa R_i}{1+\kappa R_i} +
\left(1-\frac{\epsilon}{\epsilon_i}\right)
\right].
\ee
\section{Partition coefficients and phase diagrams}
\label{partition}
\begin{figure}[t]
\begin{center}
\includegraphics[height=5cm]{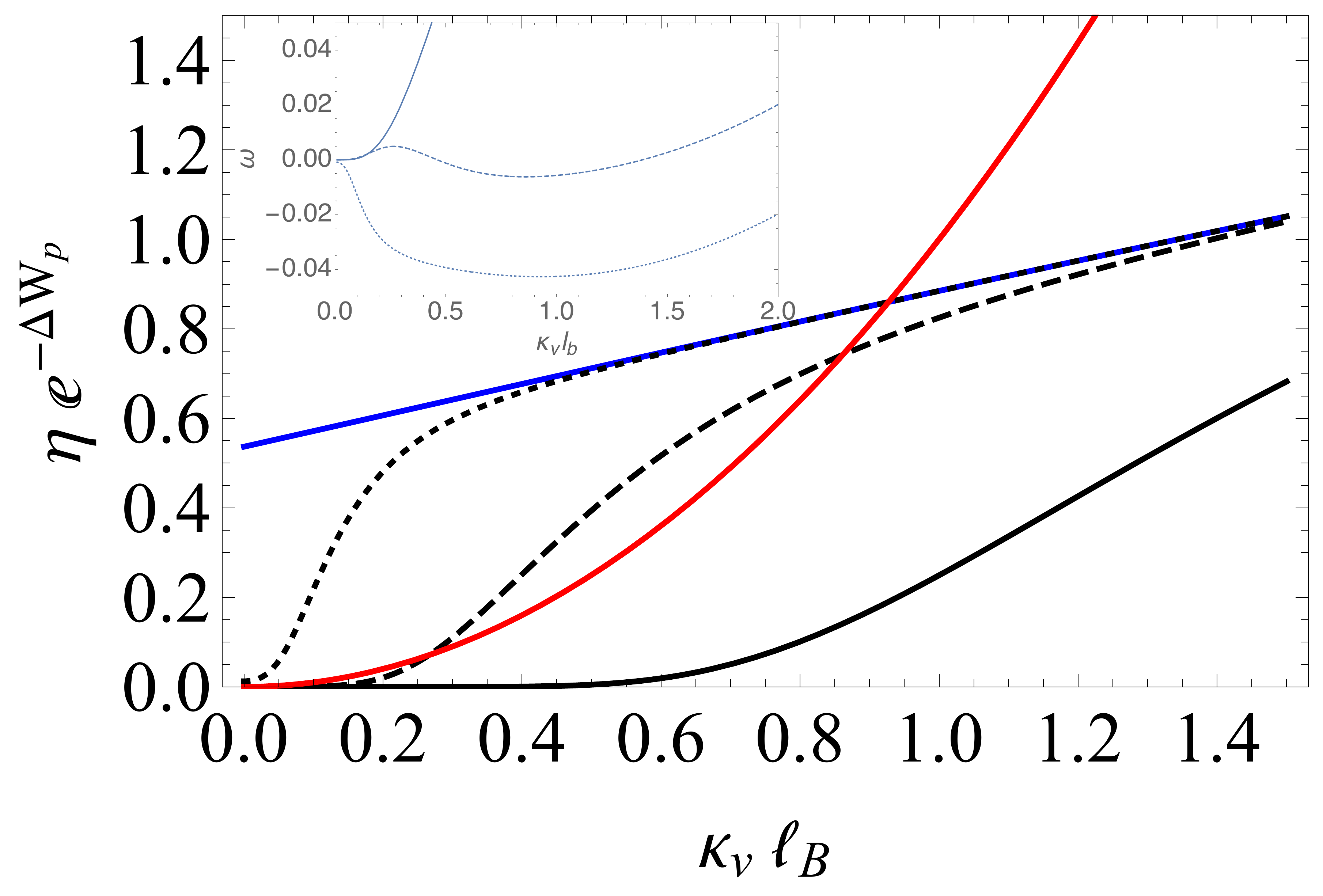}\hspace{1cm}
\includegraphics[height=5cm]{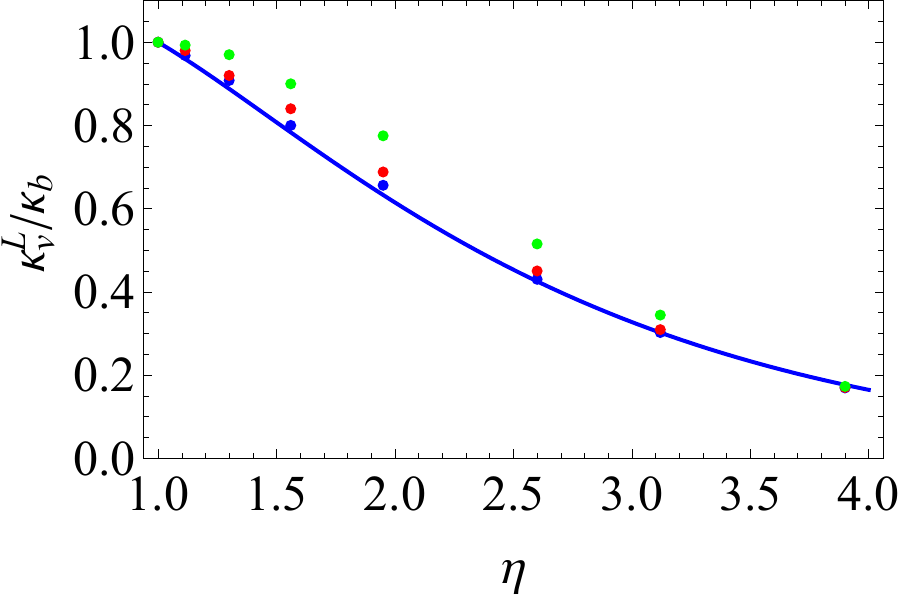}\\
(a)\hspace{7cm}(b)
\caption{(a)~Graphical solution of the variational equation \eq{eq:kappaminim2} for $\kappa_b\ell_B=1$, $R_i=0.3\ell_B$, $\epsilon=60$, $\epsilon_m=2$ and three different radii, $R/\ell_B=1$ (solid line), 2 (dashed line), 5 (dotted line). The black curves correspond to $\eta e^{- \Delta W_{\rm p}(\kappa_v)}$ and the red curve to $(\kappa_v/\kappa_b)^2$. For $R=\ell_B$ (respectively $R=5\ell_B$) the solution corresponds to the vapour (resp. liquid) phase.  The case $R=2\ell_B$ is in the liquid phase but with a metastable vapor state. The blue curve corresponds to the asymptotic behavior $\eta e^{-W_{\rm DH}(\kappa_v)-W_{\rm Born}}$. Inset: Associated grand-potential $\omega$, the minima of which correspond to the stable or metastable solutions. (b)~Numerical solutions (symbols) corresponding to the intersection between the red and blue curves in~(a) versus $\eta=\epsilon_b/\epsilon$ for $\kappa_b\ell_B=0.5$ (blue), 1 (red) and 2 (green). The curve corresponds to the approximate solution given in \eq{approx}.}
\label{f2}
\end{center}
\end{figure}

We first focus on the case where the pore surface is neutral $\sigma=0$ (and therefore, according to \eq{electro}, $\Phi_0=0$, because we are considering only symmetric electrolytes). The partition coefficient is the same for coions and counterions, $k= e^{-\Delta W_{\rm p}(\kappa_v)}$. By minimizing the grand-potential $\omega(\kappa_v)$, we find one stable physical solution (ionic fluid phase) or two physical (vapor and liquid) solutions for $\kappa_v$ (stable or metastable) [and one unstable (unphysical) solution], corresponding to a first order phase transition, depending on the values of $\rho_b$ (or equivalently $\kappa_b^2$) and $R$. The variational equation \eq{eq:kappaminim} can be rewritten
\be
\frac{\kappa_v^2}{\kappa_b^2}= \eta e^{- \Delta W_{\rm p}(\kappa_v)}.
\label{eq:kappaminim2}
\ee
When there are three solutions, the (stable or metastable) solution where $\kappa_v\gtrsim0$ corresponds to an ionic vapor phase with almost no ions entering the nanopore. The second (stable or metastable) solution is obtained for $\kappa_v \simeq \kappa_b$ and corresponds to an ionic liquid phase. The solution in between corresponds to a maximum of $\omega(\kappa_v)$ and is therefore unstable. The graphical determination of the solutions is illustrated in \fig{f2}(a) for $\kappa_b\ell_B=1$, $R_i=0.3\ell_B$, $\epsilon=60$, $\epsilon_m=2$ and three different radii, $R/\ell_B=1, 2, 5$. The transition occurs as soon as the value of $\Delta W_{\rm p}$ is large at small $\kappa$ and its variations are abrupt enough, i.e. when the dielectric contribution $W_{\rm diel}$ is large enough.

Although it is impossible to obtain an analytical expression for the solutions of \eq{eq:kappaminim2}, one can give a simple estimate in the limit of large pore radius $R$ for which $W_{\rm conf}$ is negligible. Indeed, in this limit the solution for the ionic liquid phase (at large $\kappa_v$) is essentially controlled by the Born and to a lesser extent DH contributions to the PMF. Graphically it corresponds to the intersection between the blue and red curves in \fig{f2}(a). Hence an analytical estimate of the solution in the liquid state, \textit{a priori} valid at large $R$, is obtained by neglecting both the confinement and DH contributions in the PMF \eq{W}, leading to
\be
\kappa_v^{\rm L} \approx \kappa_b \sqrt{\eta} e^{-W_{\rm Born}/2}=\kappa_b\sqrt{\eta}\exp\left[-\frac{z^2\ell_B}{4R_i}(\eta-1)\right]
\label{approx}
\ee
One can check in \fig{f2}(b) that this expression is in good agreement with the numerical solutions obtained by neglecting $W_{\rm conf}$ in \eq{eq:kappaminim2}. It is slightly less good for large $\kappa_b$ since the DH contribution to the PMF becomes more important. In particular, one notices that $\kappa_v^{\rm L}$ decreases when $\eta$ increases, since the Born effect, $e^{-W_{\rm Born}/2}$, overwhelms the factor $\sqrt{\eta}$ coming from the increase of the Bjerrum length.
\begin{figure}[t]
\begin{center}
\includegraphics[height=5cm]{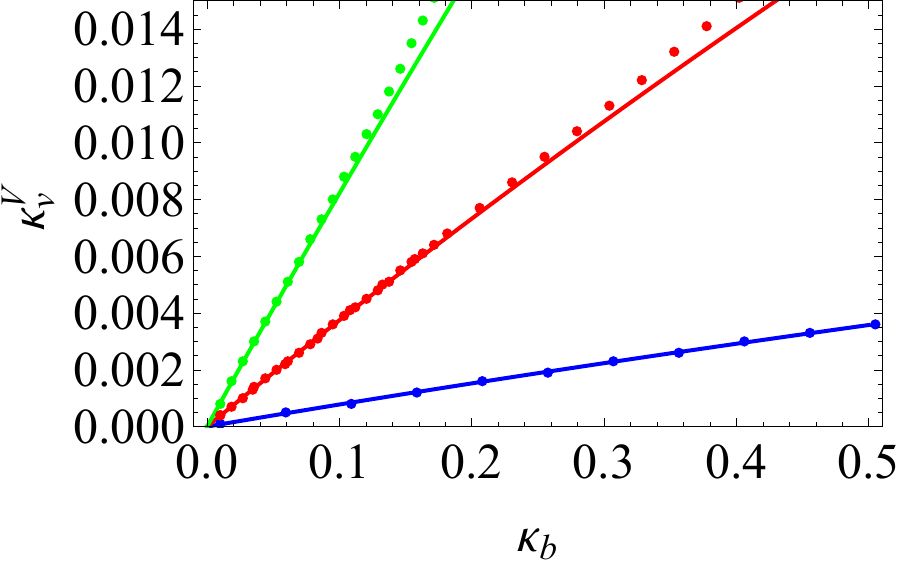}
\caption{Numerical solutions (symbols) of the variational equation  \eq{eq:kappaminim2} in the  ionic vapor phase versus $\kappa_b$ for $R/\ell_B=2$ (blue), 3 (red) and 4 (green) ($R_i=0.3\ell_B$, $\epsilon=60$ and $\epsilon_m=2$). The curves corresponds to the approximate solution given in \eq{approx2}.}
\label{f3}
\end{center}
\end{figure}

Knowing that the ionic vapor phase is obtained for low values of $\kappa_v$, a good and simple estimate of the solution to \eq{eq:kappaminim2} in this phase, $\kappa_v^{\rm V}$, is obtained by taking the limit $\kappa_v\to 0$ on the rhs. of \eq{eq:kappaminim2}. This approximation allows one to make a connection between this limit of the variational approach  and  the salt-free pore result previously obtained in Section II (Eq. 11).
One then has
\be
\kappa_v^{\rm V} \approx \kappa_b \sqrt{\eta} e^{-\Delta W_{\rm p}(0)/2}
\label{approx2}
\ee
where
\be
\Delta W_{\rm p}(0)=\frac{z^2\ell_B}{2} \left(\frac{\epsilon_b/\epsilon_m-\eta}{R}+ \frac{\eta-1}{R_i} + \frac{\kappa_b}{1+ \kappa_b R_i}\right)
\label{Wp0}
\ee
which are, respectively, the dielectric and Born contributions for an empty pore (see~\eq{dWp}) minus the classical DH chemical potential in the bulk.
We compare this result with the numerical solution in \fig{f3}. The agreement is extremely good at low $\kappa_b$. One could in principle improve the approximate solution \eq{approx2} by expanding $\Delta W_{\rm p}(\kappa_v)$ to order 2 in $\kappa_v$, but this leads to a much complicated expression. Interestingly, one notices that the ratio between $\kappa_v^{\rm V}$ and $\kappa_v^{\rm L}$ is essentially controlled by the first dielectric contribution of \eq{Wp0}, i.e. the nanopore radius $R$ and the dielectric jump $\epsilon_m^{-1}-\epsilon^{-1}$.\\

\begin{figure}[t]
\begin{center}
\includegraphics[width=0.47\textwidth]{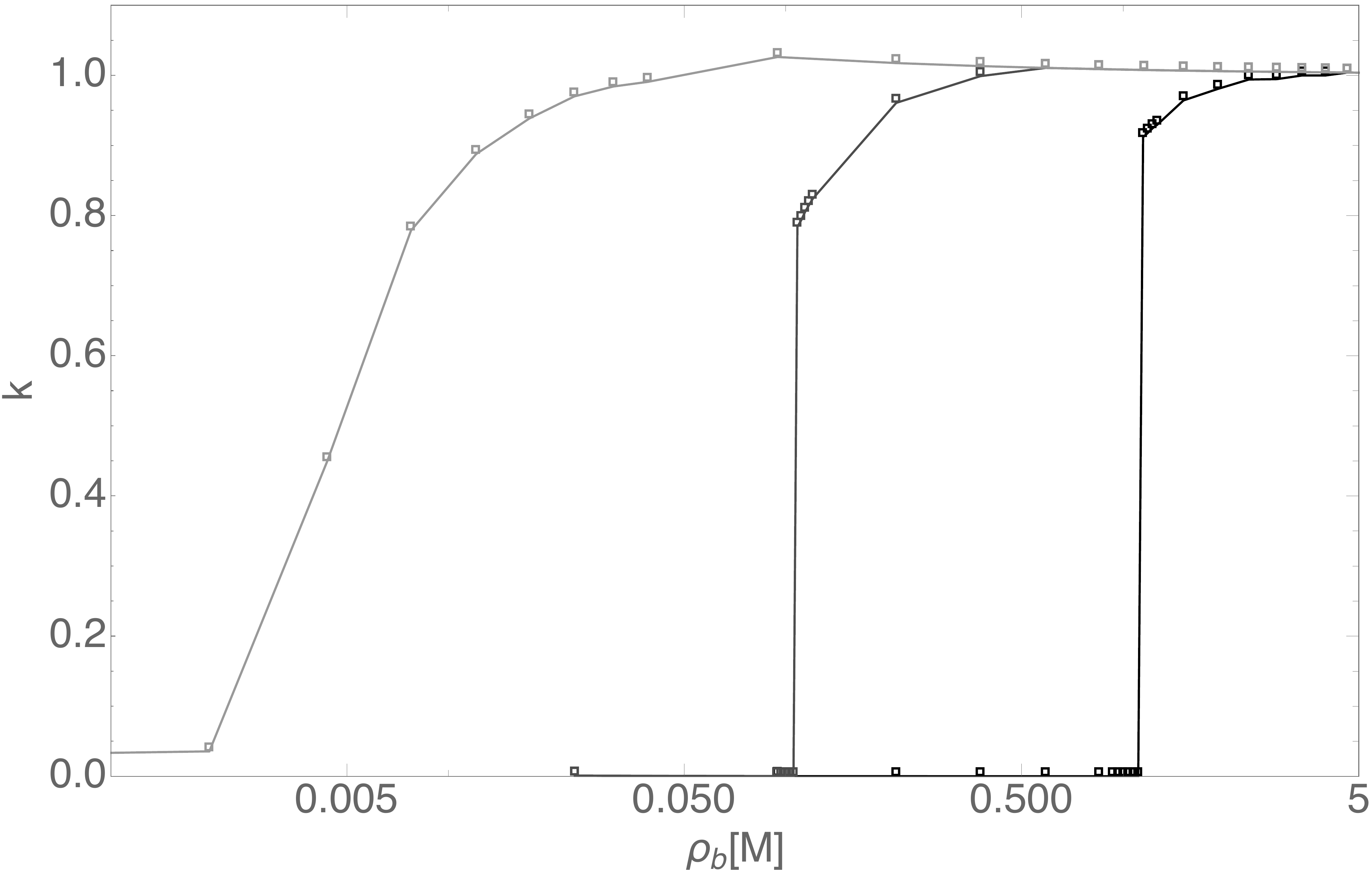}
\includegraphics[width=0.48\textwidth]{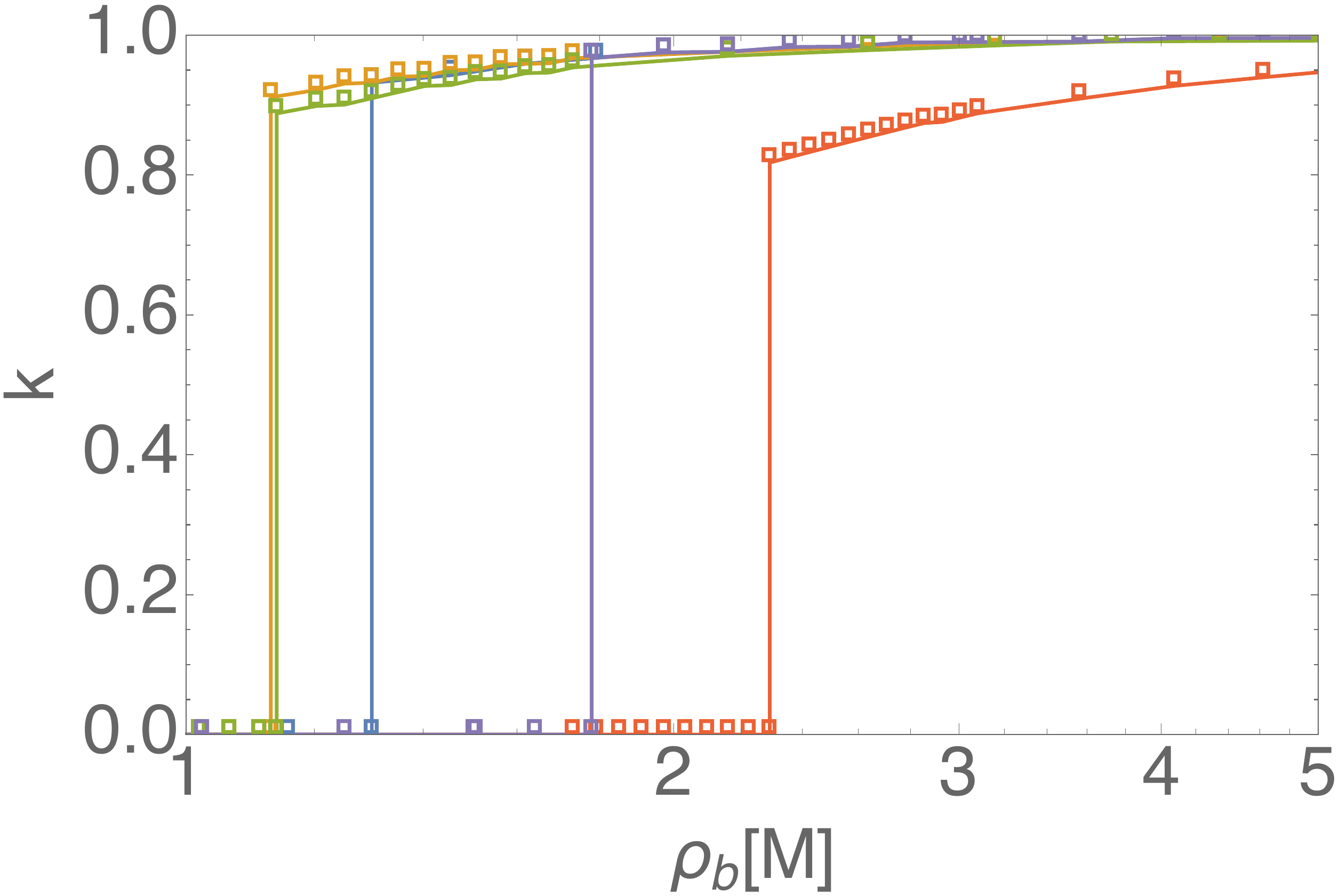}\hspace{4cm}(a)\hspace{8.5cm}(b)\hfill\\
\includegraphics[width=0.47\textwidth]{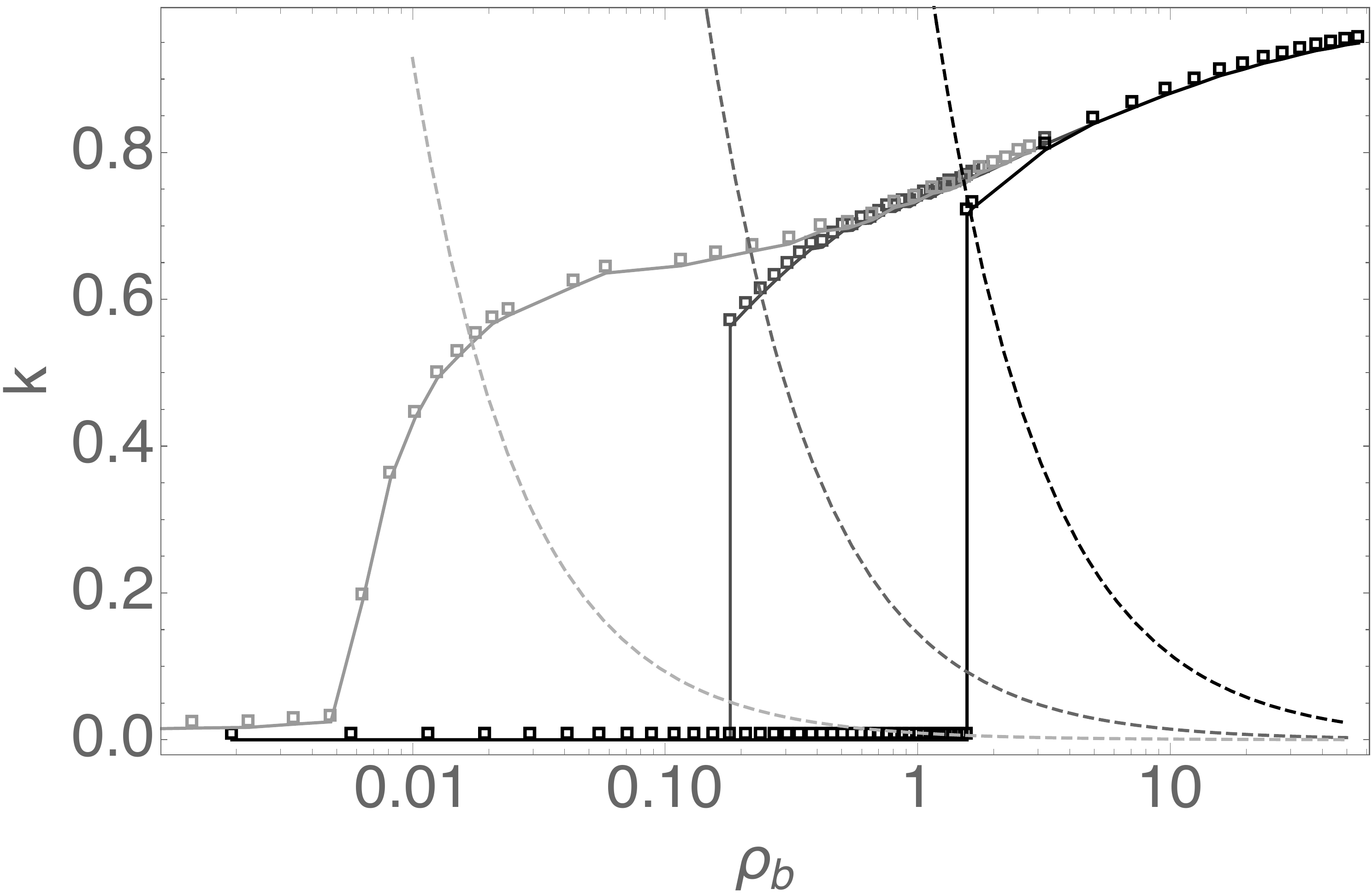}
\includegraphics[width=0.46\textwidth]{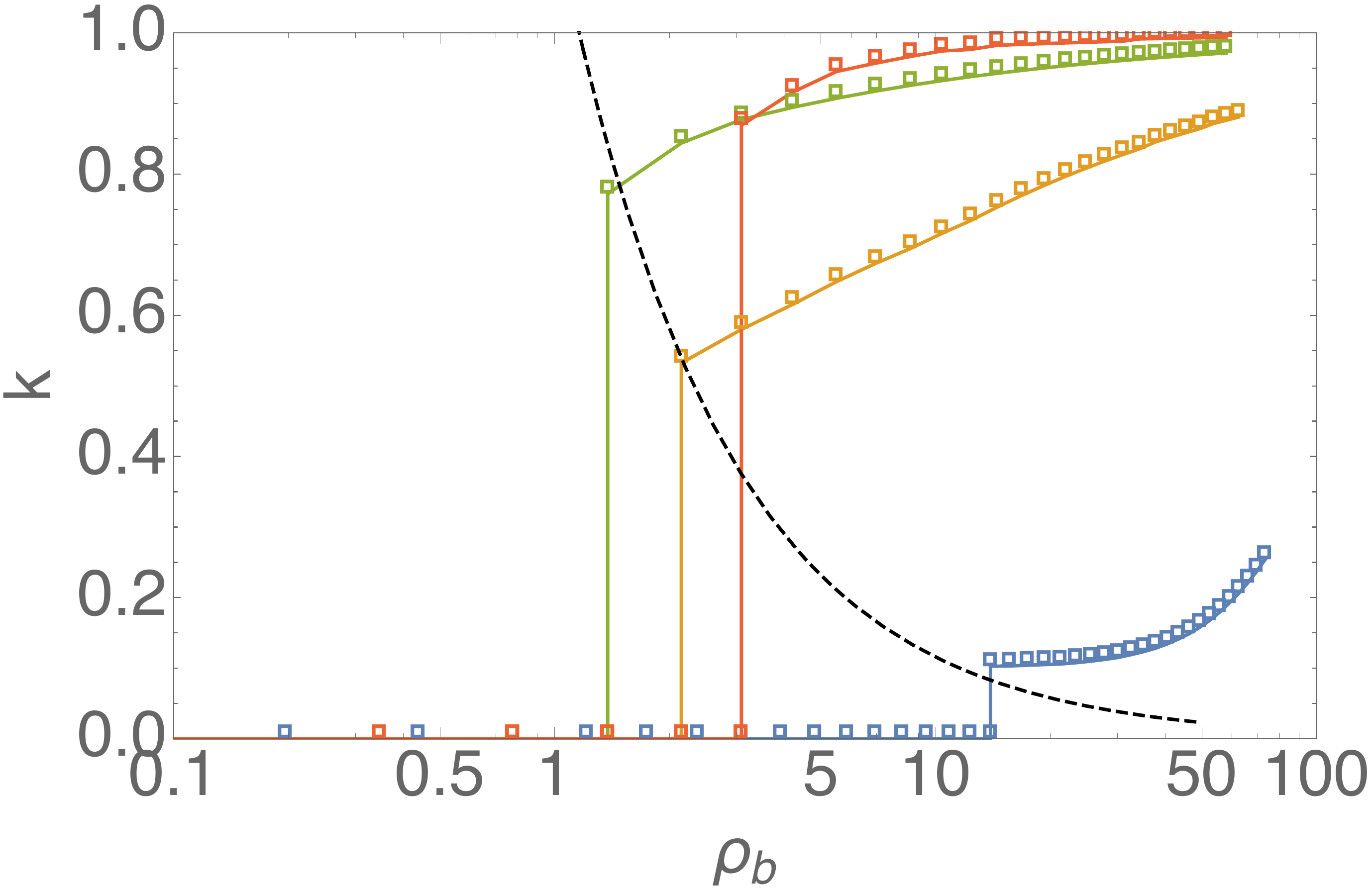}\hspace{4cm}(c)\hspace{8.5cm}(d)\hfill
\caption{(a) Ionic partition coefficient $k=\rho/\rho_b=\kappa_v^2/(\eta\kappa_b^2)$ vs the bulk concentration $\rho_b$ (log-linear plot) for various neutral pore radii ($R/\ell_B=1,2,5$ from right to left), $\epsilon = \epsilon_b=78$ and $\epsilon_m=2$. The transition from the ionic vapor phase ($k\simeq 0$) and the ionic liquid phase ($k\simeq 1$) occurs for small radii and disappears for larger pores. The effective ionic radius has been fixed at $R_i=0.3 \ell_B$. (b)~Same figure for different values of the effective ionic radius $R_i/\ell_B=0$ (purple), 0.1 (blue), 0.2 (orange), 0.4 (green), 0.8 (red) ($R=\ell_B$ and $\epsilon=\epsilon_b$). (c)~Same as (a) for $\epsilon=60$ ($\eta=1.3$) (corresponding to \fig{f2}). One clearly observes the effect of the Born self-energy that increases the critical $\rho_b$ and decreases the limiting value of $k$ in the liquid state. (d)~Same as~(b) for $\epsilon=60$ ($\eta=1.3$) and $R_i/\ell_B=0.1$ (blue), 0.2 (orange), 0.4 (green), 0.8 (red). The dashed lines correspond to the case of a single ion on average in the pore.}
\label{k_rhob}
\end{center}
\end{figure}

The partition coefficient $k$ is shown in \fig{k_rhob}a for nanopore radius $R=1, 2$ and $5\ell_B$ and $\eta=1$ (without Born exclusion). For the smallest  radii a discontinuous increase of $k$ occurs at coexistence values of the bulk concentration:  for $R=\ell_B$ the coexistence value is $\rho_b^{\rm coex} \simeq1.2$~M
and for $R=2\ell_B$, it decreases to $\rho_b^{\rm coex}\simeq 0.11$~M. This discontinuous transition disappears when $R=5\ell_B$, leading to a smooth increase of $k(\rho_b)$ with $\rho_b$.
This transition has already been predicted theoretically without Born exclusion for point-like~\cite{PRL2010,JCP2011} and finite-sized ions~\cite{JCP2016} in cylinders. As already noticed in~\cite{PRE2010} the mid-point approximation used here leads to larger values of $\rho$ in the bulk phase close to $\rho_b$, whereas excluded volume interactions must be properly included to obtain $\rho\simeq\rho_b$ without this approximation, as shown in \cite{JCP2016}.

To illustrate the role of the  finite ionic size in $W_{\rm diel}$ and $W_{\rm sol}$, we present in \fig{k_rhob}b  the variation of $k(\rho_b)$ for various ionic radii and $\eta=1$: $R_i=0$ (point-like ions), 0.1, 0.2, 0.4 and $0.8\ell_B$ for $R=\ell_B$. In all cases, one still observes the discontinuous transition but the variation of the coexistence bulk concentration, $\rho_b^{\rm coex}$, with $R_i$ is non-monotonous: it first decreases slightly when $R_i$ increases up to $R_i=0.2\ell_B$ and then increases, reaching a larger value than for point-like ions for $R_i=0.8\ell_B$. This is directly related to the variation of the dielectric contribution to the PMF, $W_{\rm diel}$, which is also non-monotonous [see \fig{f1}], decreasing when $R_i$ increases up to approximatively $R_i\sim R/2$ and then increasing again due to the term in $\exp[2\kappa(R_i-R)]$ in the first term of $W$ in \eq{W}.

The role of the Born self-energy in the PMF, $W_{\rm Born}$, is highlighted in \fig{k_rhob}(c) for $\eta\neq1$ and in \fig{k_rhob}~(d) for various values of $R_i$. The partition coefficients shown in these two  figures have a lower saturation value, controlled by $\exp(-W_{\rm Born})$ which is independent of $\kappa_v$. Hence, even in the liquid state, the Born solvation energy decreases the concentration in the pore. This is a direct consequence of the result that the Born contribution dominates the PMF in the liquid state (see \fig{f1}). Moreover the transition is shifted to higher bulk critical values, especially for small ions, due to the factor $1/R_i$ in $W_{\rm Born}$.
The dashed lines correspond to the case of a single ion in the pore on average, fixed by $k=(\frac43\pi R^3\rho_b)^{-1}$. One clearly see that it crosses the transition for all $R$ [\fig{k_rhob}(c)] and $R_i$ [\fig{k_rhob}(d)]. Hence the vapor state corresponds to less than one ion in the pore on average. As in \fig{k_rhob}(b), \fig{k_rhob}(d) shows a non-monotonous variation of the critical bulk concentration with $R_i$.\\
\begin{figure}[t]
\begin{center}
\includegraphics[width=0.48\textwidth]{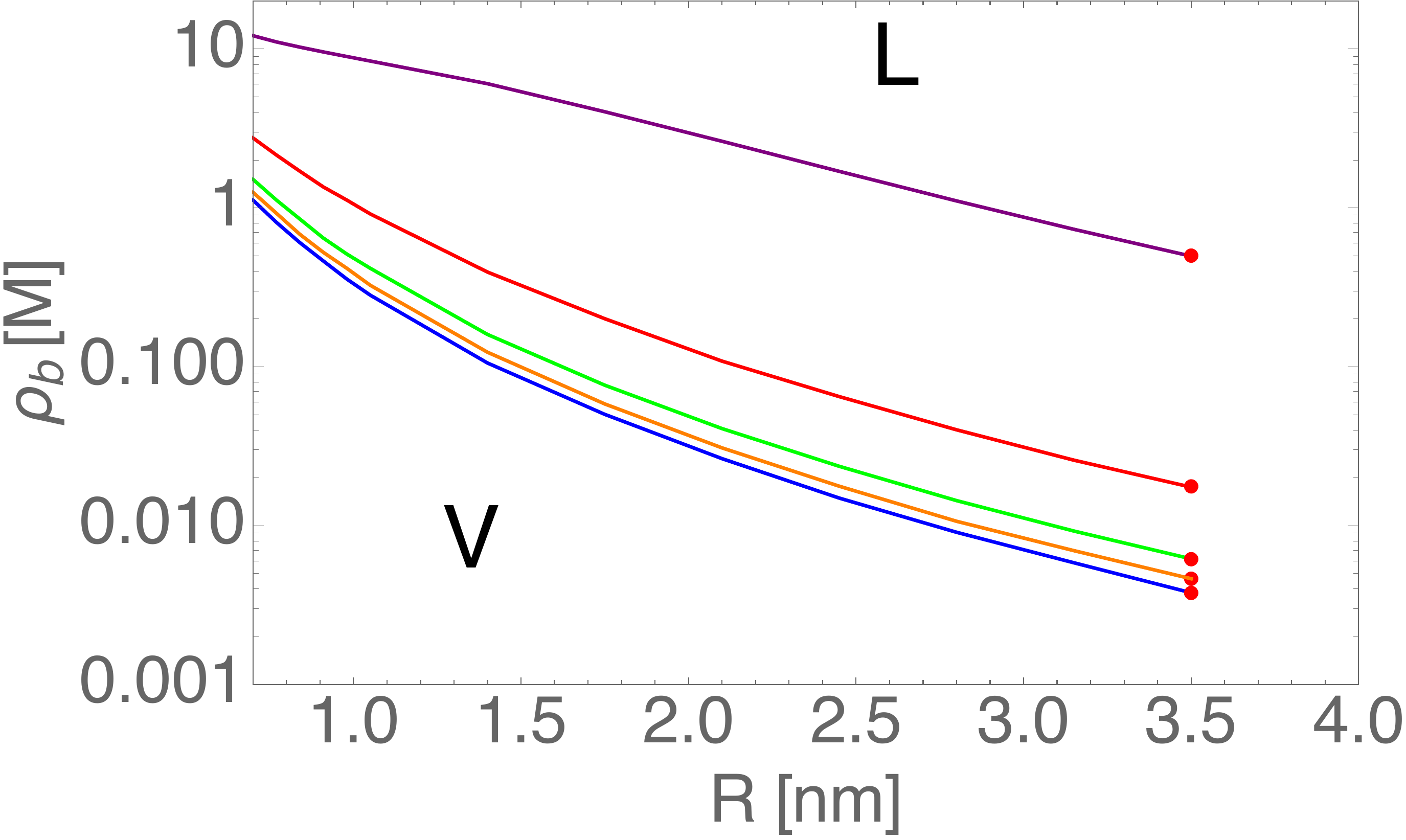}
\includegraphics[width=0.46\textwidth]{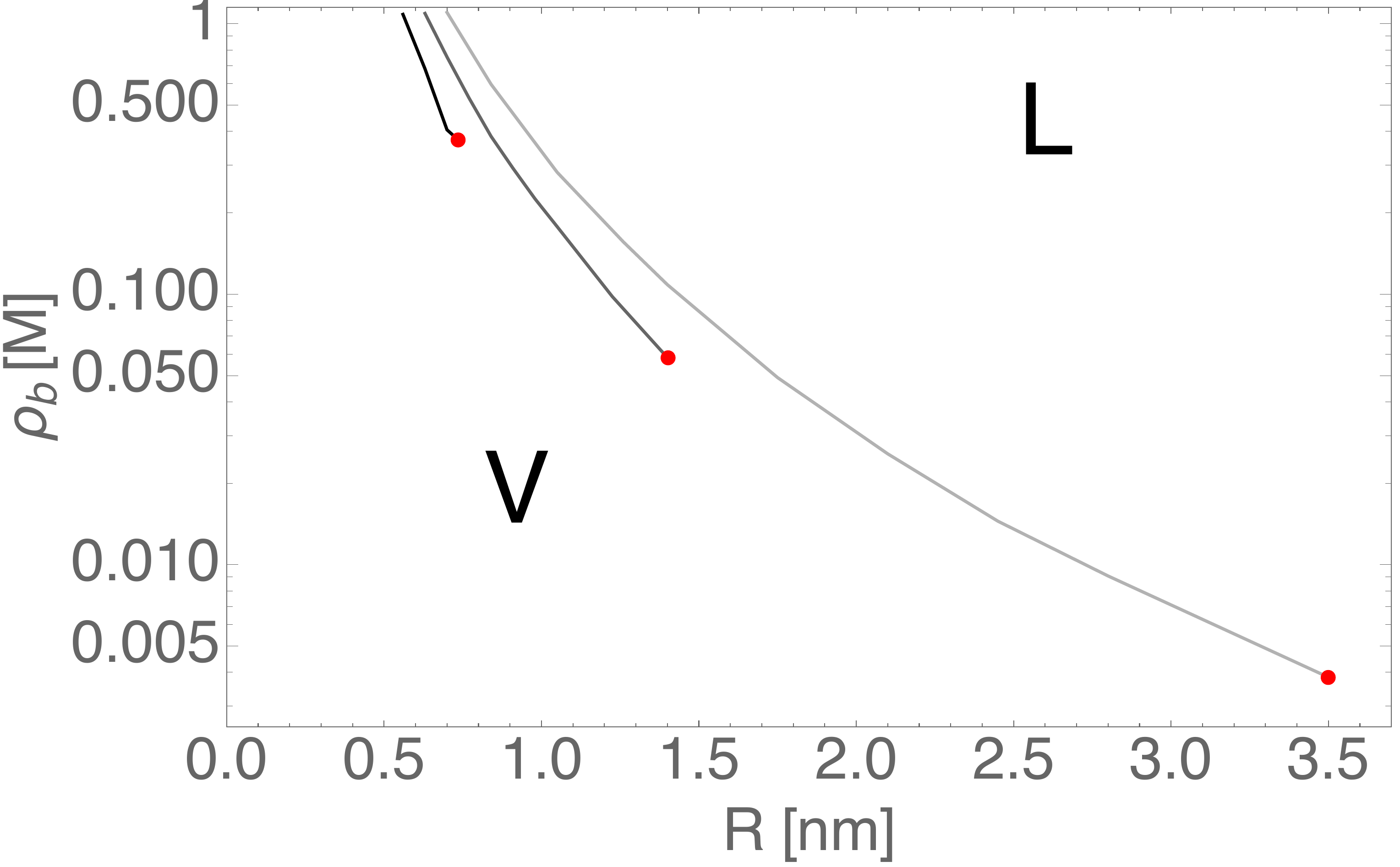}
\hspace{4cm}(a)\hspace{8.5cm}(b)\hfill
\caption{(a) Phase diagram (linear-log plot) in the  pore radius $R$ vs. bulk concentration $\rho_b$ plane for 3 different nanopore dielectric constants ($\epsilon=20,40,60,70, 78$ from top to bottom) for $R_i=0.3\ell_B$ and $\epsilon_m=2$. The left bottom (respectively right top) corner corresponds to the ionic vapor (resp. liquid) phase, and the critical point $(R^*,\rho_b^*)$ is shown in red. (b) Same figure as in (a) but for $\epsilon_m=2,5,10$ from right to left ($\epsilon=60$ and $R_i=0.1\ell_B$, linear-log plot).}
\label{PD1}
\end{center}
\end{figure}
By plotting this coexistence concentration $\rho_b^{\rm coex}$ corresponding to the first order transition as a function of the nanopore radius, we can construct the phase diagram shown in \fig{PD1}(a) for four different values of the confined water dielectric constant, $\epsilon$, and fixed membrane one, $\epsilon_m=2$. For large $\rho_b$ and $R$ the nanopore is in the ionic liquid state, and below the coexistence curve, for low $\rho_b$ and $R$ it is in the ionic vapor one. Clearly the coexistence line moves to higher $\rho_b$ values when $\epsilon$ decreases (i.e. $\eta$ increases). Indeed, the dielectric contribution to the PMF $W_{\rm diel}$ is proportional to $\eta$ for fixed $\kappa_v$, which thus favors the ionic vapor state. Interestingly, the critical radius $R^*$ (in red) does not change with $\epsilon$. This is probably due to the result that the critical point is fixed by the abrupt decrease of
$\Delta W_{\rm p}(\kappa_v)$, which is almost independent of the Born self-energy [as shown in the inset of \fig{f1}(b)].

In \fig{PD1}(b) we illustrate the role of the membrane dielectric constant $\epsilon_m$, by changing its value from 2 to 10 for a fixed $\epsilon=60$. Increasing $\epsilon_m$ moves the coexistence line to the left and favors the liquid state, but the critical point also moves to lower radii $R^*$ and larger critical bulk concentrations $\rho_b^*$.

\begin{figure}[t]
\begin{center}
\includegraphics[width=0.49\textwidth]{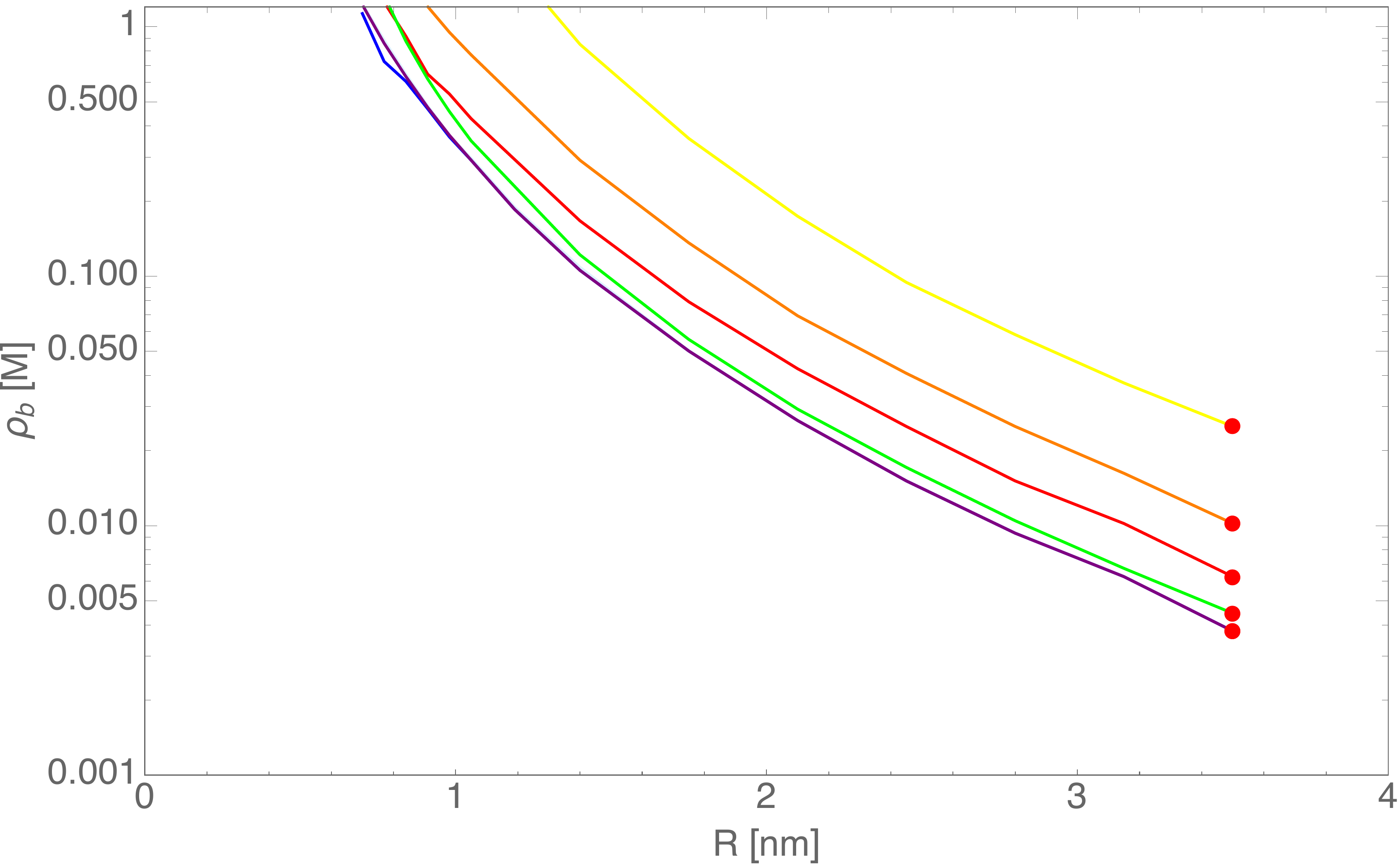}
\caption{Same as Fig.~\ref{PD1} for three different ion radii ($\epsilon_m=2$): for $\epsilon=78$  the three curves are almost superimposed [$R_i/\ell_B=0$ (light blue), 0.2 (blue), and 0.5 (purple)], whereas they are shifted to larger $\rho_b$ values when $R_i$ decreases for $\epsilon=60$ [$R_i/\ell_B=0.1$ (yellow), 0.2 (orange), 0.5 (red), and 0.8 (green)].}
\label{PD2}
\end{center}
\end{figure}

We illustrate the influence of the ion radius $R_i$ in \fig{PD2} using $R_i=0,0.1,0.2,0.5$ and $0.8\ell_B$ for (a)~$\epsilon=\epsilon_b$ and (b)~$\epsilon= 60$.
The fact that the curves are almost superimposed in case~(a) confirms that $R_i$ plays an essential role only in the Born exclusion (last term on the RHS of \eq{W}). Indeed the terms that depend on $\kappa R_i\ll1$ do not influence much $W_{\rm diel}$ and $W_{\rm sol}$. On the contrary, for $\eta>1$, \fig{PD2} shows that the liquid state is favored when $R_i$ increases, owing to the decrease in $1/R_i$ in $W_{\rm Born}$, leading to a decrease in Born exclusion. The non-monotonous behavior observed in \fig{k_rhob}(d) (for $R=\ell_B=0.7$~nm) occurs only for small nanopore radius $R$. In any case the shift of the coexistence curve remains small.

\begin{figure}[t]
\begin{center}
\includegraphics[width=0.47\textwidth]{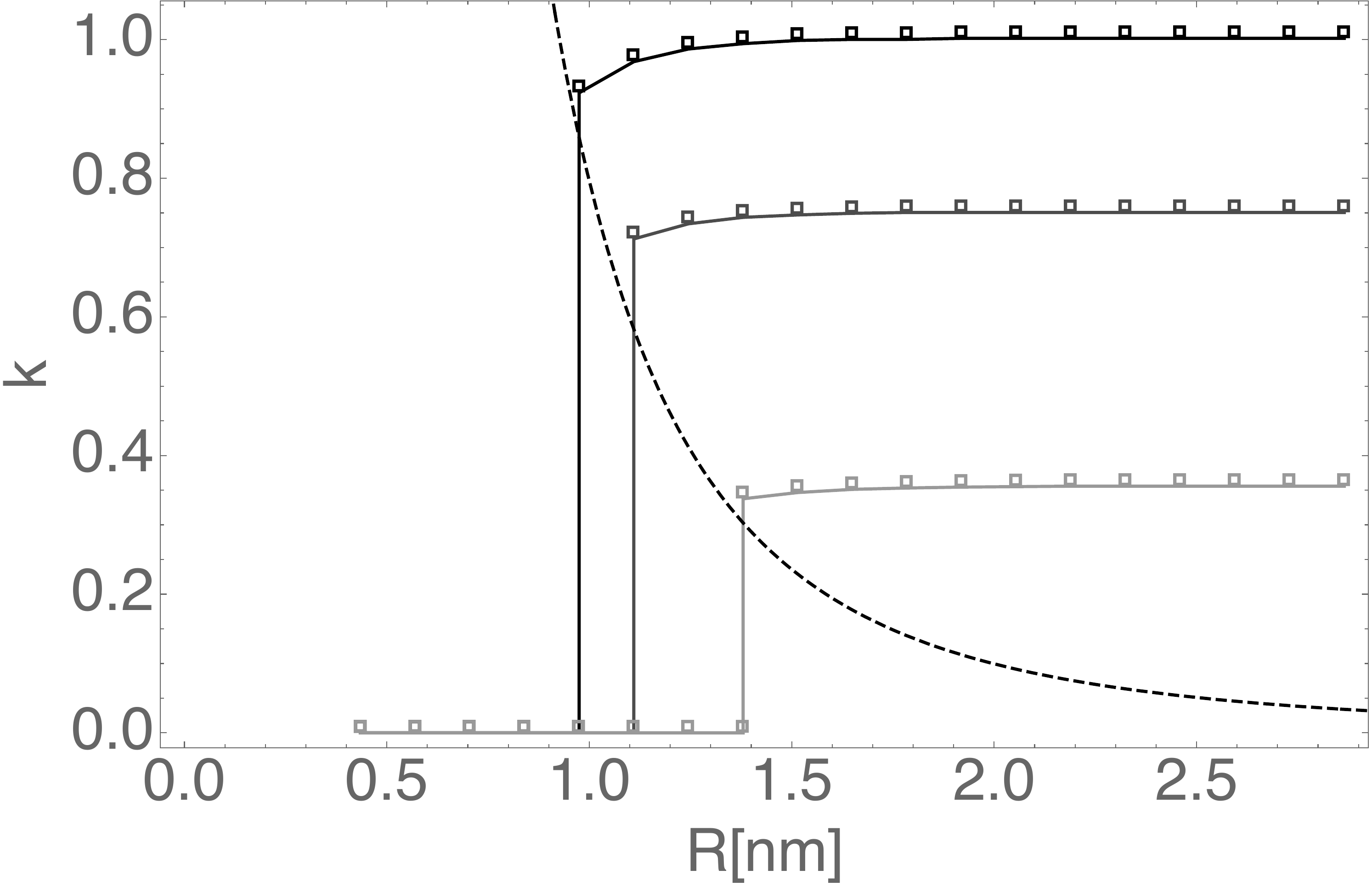}
\includegraphics[width=0.49\textwidth]{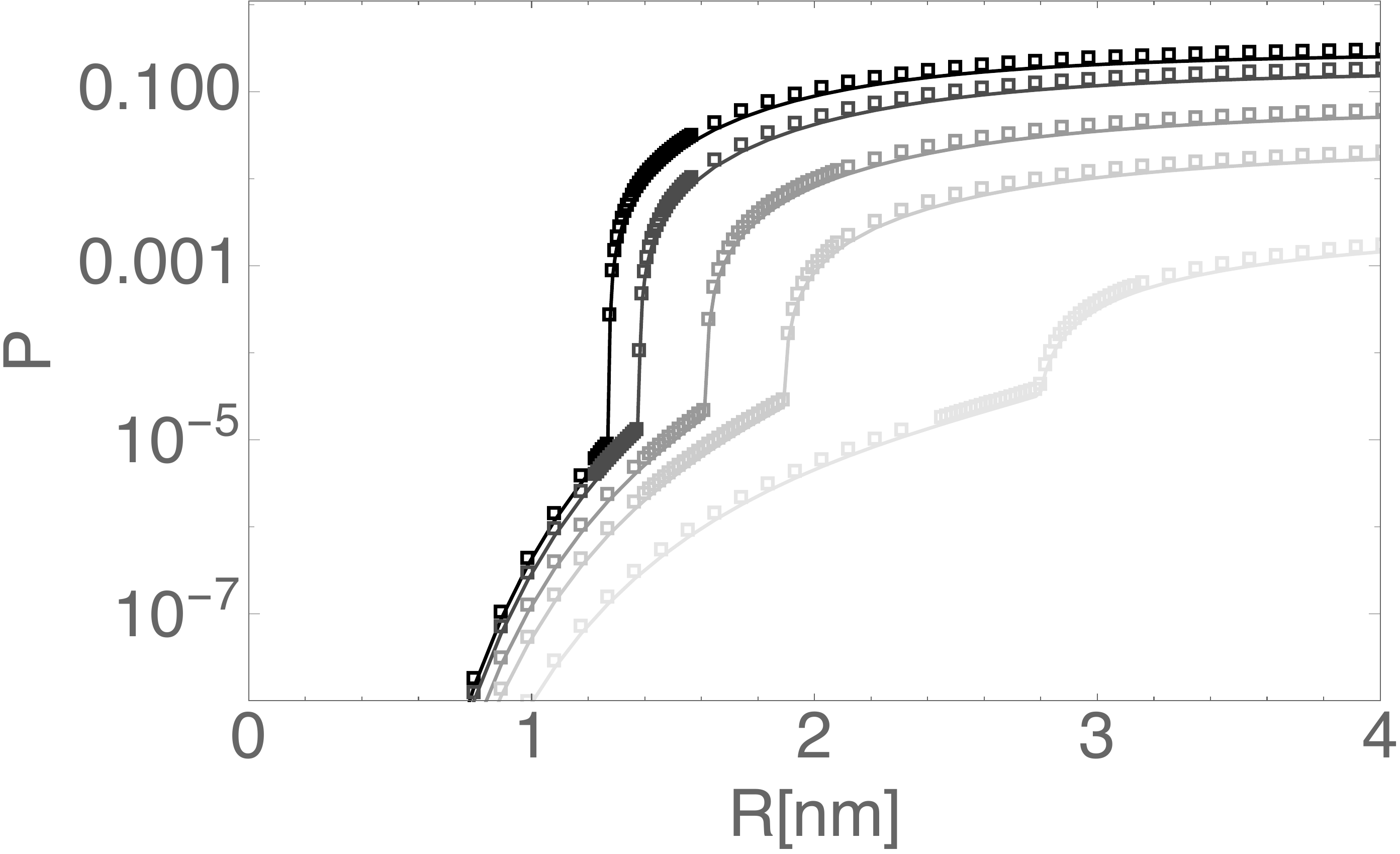}
\hspace{4cm}(a)\hspace{8.5cm}(b)\hfill
\caption{Influence of the pore radius $R$ (for $\rho_b=0.5$~mol/L, $R_i=0.3\ell_B$) on (a)~ the partition coefficient $k$ (the dashed line corresponds to the case of a single ion on average in the pore)  for $\epsilon=\epsilon_b,60,40$ from left to right. (b)~Total osmotic pressure $p=-k_BT \omega$ (in units of $k_BT/\ell_B^3\simeq 1.1\times 10^7$~Pa at room temperature) versus $R$ for $\epsilon=\epsilon_b,60,40,30,20$ from left to right.}
\label{effectR}
\end{center}
\end{figure}

We study the influence of the pore radius $R$ on the partition coefficient $k$ and the total osmotic pressure in the pore $p=-k_BT\omega$ by inserting $\kappa_v$ after minimization for $\rho_b=0.5$~mol/L (see \fig{effectR}). Similarly to \fig{k_rhob}(a) and (c), $k$ abruptly increases from the ionic vapor  state at low radius to the liquid one for large radius with a critical radius of $R\simeq 0.9$~nm for $\eta=1$, $\simeq 1$~nm for $\eta=1.3$ ($\epsilon=60$) and $\simeq 1.2$~nm for $\eta=1.95$ ($\epsilon=40$). The saturation value for $k$ is quickly reached and decreases from 1 to 0.4 when $\eta$ increases. It corresponds to the decrease of $\kappa_v^L$ given in \eq{approx} since $k\approx(\kappa_v^L)^2/(\eta \kappa_b^2)= e^{-W_{\rm Born}}$. Hence, in the liquid state and at fixed $R$, the concentration in the pore $\rho$ decreases when $\eta = \epsilon_b/\epsilon$ increases.
The signature of the transition also appears in \fig{effectR}(b) where the pressure increases by 3 orders of magnitude at the transition for $\eta=1$. When $\eta$ increases the transition occurs at larger radius and the jump is smaller, of one order of magnitude for $\epsilon=20$.

In the liquid state and for large $\rho_b$ or large $R$, the pressure is given by the grand-potential, $\beta p=-\omega(\kappa_v^L)$, by assuming that the pore term (which depends on $W_{\rm conf}$) is negligible. In this limit the pressure can therefore be approximated by the usual bulk DH result (see, e.g.,~\cite{McQuarrie}), but with a salt concentration and dielectric constant appropriate for a nanopore and in general different from the real bulk values:
\be
\beta p_b=2\rho_b\ e^{-W_{\rm Born}} -\frac1{8\pi R_i^3} \left[2 \kappa_v^L R_i-2 \ln\left(1+\kappa_v^L R_i\right)-\frac{(\kappa_v^L R_i)^2}{1+\kappa_v^L R_i}\right],
\ee
with $\kappa_v$ approximated by $\kappa_v^L$, given in \eq{approx}, which is a very accurate for large $R$.
This equivalent bulk pressure decreases when $\eta$ increases essentially due to the first ideal term, since the concentration in the pore $\rho$ decreases when $\eta$ increases. It is roughly $0.18 k_BT/\ell_B^3\simeq 20$~bar (respectively $0.06k_BT/\ell_B^3$) for $R_i=0.3\ell_B$, $\rho_b=0.5$~mol/L and $\eta=1$ (resp. $\eta=1.95$). It indeed corresponds to the limiting values for large $R$ of the black and light grey curves respectively observed in \fig{effectR}b.
Hence the pressure is a good observable to study the ionic LV transition and the influence of $\eta$ on the critical radius.\\

\begin{figure}[t]
\begin{center}
\includegraphics[width=0.49\textwidth]{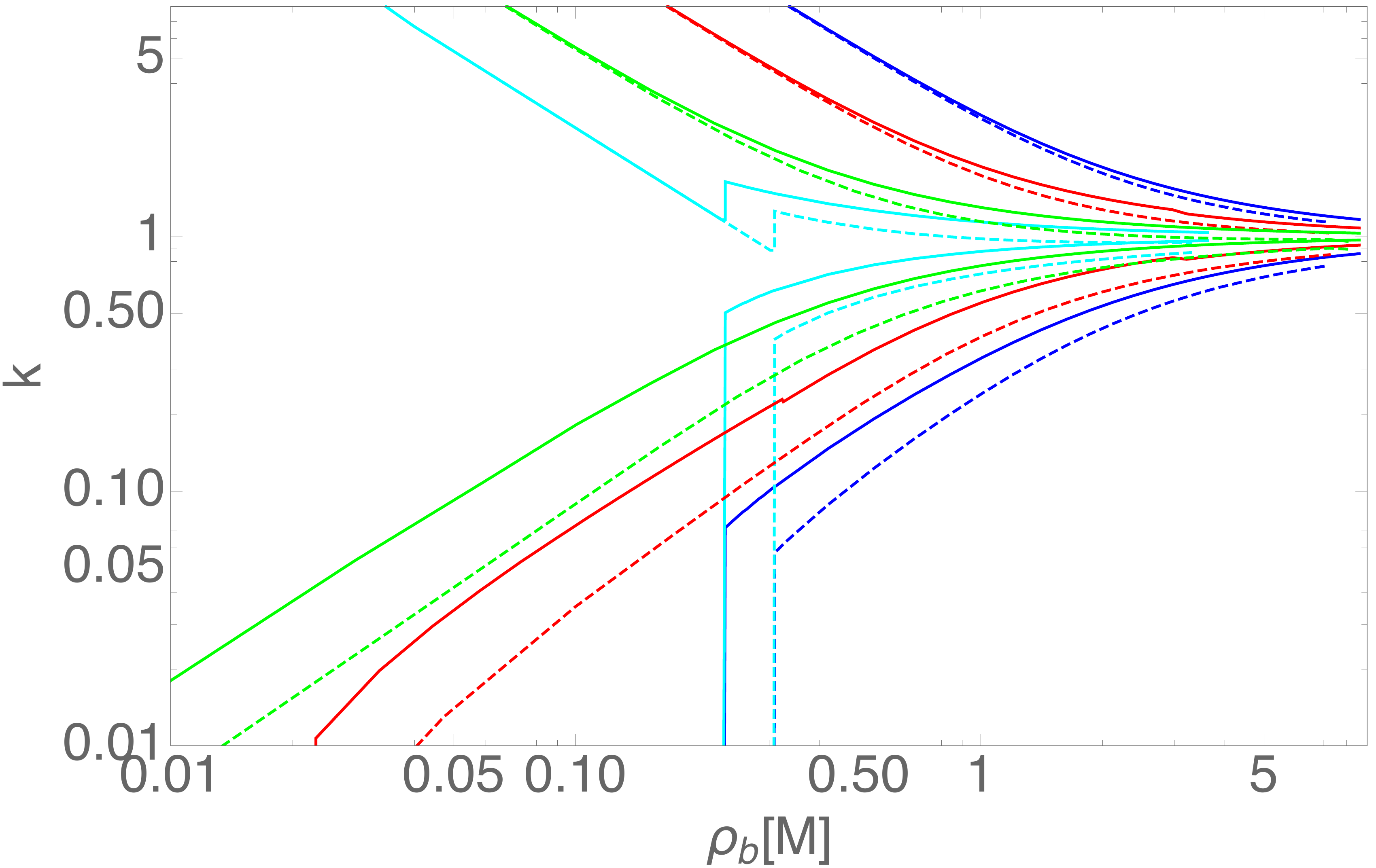}
\caption{Partition coefficient of counterions $k_+(\rho_b)$ (top) and co-ions $k_-(\rho_b)$ (bottom) for various pore radii with $\epsilon=78$ (solid lines) and $\epsilon=60$ (dashed lines) for a charged nanopore with $\sigma= -0.02$~C/m$^2$  ($R_i=0.3\ell_B$): $R=\ell_B$ (blue), $R=2\ell_B$ (red), and $R=5\ell_B$ (green). The transition is also visible on $k_+$ for a smaller surface charge $\sigma= -0.002$~C/m$^2$ (cyan).}
\label{charge}
\end{center}
\end{figure}

Finally, we consider the case of a slightly charged pore with $\sigma=-0.02$~C/m$^2$.
We plot in \fig{charge} the partition coefficient of counterions $k_+(\rho_b)$ and co-ions $k_-(\rho_b)$ for $R/\ell_B=1,2$, and 5 and for $\epsilon=78$ and 60. As expected for this surface charge polarity, $k_+>1$ and $k_-<1$. Moreover, the transition remains visible on $k_-$, but not on $k_+$. Finally, as also expected, the decrease of $\epsilon$ from $\epsilon_b$ to 60 decreases both $k_+$ and $k_-$ in the liquid phase, due to Born exclusion. For low $\rho_b$, $k_-\simeq 0$ and following \eq{electro}, one has $k_+\simeq3|\sigma|/(eR\rho_b)$ which corresponds to the asymptotic lines at low $\rho_b$ observed in \fig{charge}.

For a small pore radius, $R=\ell_B$, and a lower value of surface charge density, $\sigma=-0.002$~C/m$^2$, which could be due for instance to a (partial) charge defect on the nanopore surface, the transition is also noticeable on $k_+$ at almost the same bulk density as for $\sigma=-0.02$~C/m$^2$, $\rho_b^{\rm coex}\simeq0.025$~M (respectively 0.03~M) for $\eta=1$ (resp. $\eta=1.3$). This is because at this $\rho_b$, one has $k_+=3|\sigma|/(R\rho_b)\simeq 0.35<1$, and a jump occurs at the transition to a value larger than 1. Hence, for low enough surface charge $|\sigma|$ and low nanopore radius $R$, the transition is still noticeable on the counter-ions partition coefficient. For larger $|\sigma|$ the system remains in the Good Co-ion Exclusion (GCE) regime, where $k_-\ll k_+$, for a wide range of reservoir concentration values $\rho_b$.

\section{Discussion and concluding remarks}
\label{discussion}

In this article, we have studied the statistical physics of an electrolyte in a spherical nanopore of radius in the nanometer range in contact with a bulk reservoir. We focused on the influence of the dielectric constant of the water confined inside the pore $\epsilon_m\leq\epsilon\leq\epsilon_b$ and the Born ionic radius $R_i$ (roughly equal to to the hard-core ion plus water radius) on the partition coefficient and the osmotic pressure.
To do so we first computed the PMF for a test ion located at the center of the pore, given in \eq{W}, and then determined variationally the Debye screening parameter $\kappa_v$ inside the pore by minimizing the variational grand-potential $\omega$ given in \eq{omega}. The partition coefficient inside the pore is governed by the three intertwined contributions to the PMF that take into account ionic correlations: the dielectric contribution, $W_{\rm diel}$, which is the only one that depends on the dielectric jump at the nanopore surface  ($\epsilon_m^{-1}-\epsilon^{-1}$), the solvation one, $W_{\rm sol}$, and the Born self-energy, $W_{\rm Born}$, which depends on the dielectric mismatch between the pore and the bulk ($\epsilon^{-1}-\epsilon_b^{-1}$), but does not depend explicitly on $\kappa_v$ and is therefore independent of the ionic concentration in the pore.

For $\epsilon_m=2$, the first order transition between a ionic vapor state (found for low bulk concentration $\rho_b$ and small pore radius $R$) to a liquid state (for larger $\rho_b$ and $R$), is induced by dielectric exclusion and has already been studied for point-like ions (and finite sized ones without Born exclusion). Thanks to the determination of the variational grand-potential integrating Born exclusion, we are able to construct the phase diagram and the coexistence curves in the plane ($R,\rho_b$) for this more general previously unstudied case.
We show that this transition
survives for $20\leq\epsilon\leq\epsilon_b=78$ for ions  of Born radius $R_i=0.3\ell_B\simeq 0.2$~nm. For lower $\epsilon$ values, the pore remains in the vapor state, with no ions entering the pore over the whole accessible range of $\rho_b$. We have shown how the Born self-energy decreases by a factor $\simeq \exp(-W_{\rm Born})$ the value of the partition coefficient $k=\rho/\rho_b$ in the liquid state. Moreover the critical radius $R^*$ does not change with $\epsilon$ (whereas $\rho_b^*$ increases when $\epsilon$ decreases)  and its value is fixed by $\epsilon_m$.
Finally we suggest that it would be insightful to study experimentally the influence of the low water confined dielectric constant by varying the pore radius $R$. In particular the transition studied here has a clear signature on the osmotic pressure inside the pore. For charged pores, the system remains in the Good Co-ion Exclusion (GCE) regime, where $k_-\ll k_+$, for a wide range of reservoir concentrations, $\rho_b$.

The application of our method presented in detail here can be applied to blue energy production using carbon nanotubes and other nanoporous systems where the external medium (membrane) has a lower dielectric constant than that of the confined electrolyte solution. In principle our approach can also be  extended to the opposite case where the membrane has a larger dielectric constant than the confined solution. Ions are then attracted towards the pore surface due to dielectric effects. This is for instance the case for the carbon  electrodes used to harvest blue energy. It is then essential to include properly in the theory the excluded volume interactions to avoid any unphysical increase of the ionic concentration near the pore surface. One way to do so would be to use the approach developed in~\cite{JCP2016}.

Although the ions are assumed to be in equilibrium in this study, it is well known that, in the linear response framework, the partition coefficients $k_{\pm}$ can be plugged into the electrokinetic coefficients to study electrokinetic transport in long nano-channels~\cite{Schoch, Yaroshchuk2000,revue_John,PRE}. Within this framework we intend in the future to study the role of the confined water dielectric constant on the experimentally measurable transport coefficients. An important question for ionic transport concerns the influence of dielectric mismatch on ion mobility, a topic already studied via simulations~\cite{Erik}.
Especially important industrial applications include membrane nanofiltration~\cite{Yaroshchuk2000,revue_John,Sym2007,Sym2009} and blue energy production using osmotic pressure gradients~\cite{Bocquet_charlaix}. The potential role and importance of ionic liquid-vapor phase transitions in these applications remain to be determined.

Several approximations have been adopted in this study. The main one concerns the mid-point approximation, which consists in assuming that the PMF of an ion located anywhere in the nanopore is equal to the one computed for an ion located at the pore center. The PMF for a point-like ion is given in the Appendix and shows increasing dielectric repulsion close to the pore wall. This effect, which forces the ions to be located close to the center, can and should be studied by numerical simulations. The second approximation is that we did not consider explicitly the hard core excluded volume interaction, the finite radius of the ions entering only in the electrostatic contribution to the PMF. We have shown in~\cite{JCP2016} that for cylindrical pores and without the mid-point approximation this excluded volume interaction increases the partition coefficient $k$ in the liquid phase roughly by a factor of 2, such that it saturates to 1 for $\rho_b>2$~M. Since in this study we observe the expected saturation to 1, one can suppose that there is a kind of beneficial  compensation between these two approximations. Moreover, we have assumed, in the case of a charged pore, and following our previous studies~\cite{PRE2010,PRL2010,JCP2011,JCP2016}, that electroneutrality is satisfied following \eq{electro} and that the surface charge does not appear explicitly in the self-energy $\Delta W_{\rm p}$. This issue, which has been questioned recently by Levy \textit{et al.}~\cite{Levy_Bazant} remains to be clarified for instance by Monte Carlo simulations. It would also be extremely interesting to study the ionic LV transition for the mesoscopic model proposed here using appropriate simulation techniques.

An open question in the theory of electrolytes concerns the relationship between the field theoretical variational method adopted here and the Splitting Field method~\cite{Lue2015} developed previously to study in an approximate way the subtle crossover between the mean-field Poisson-Boltzmann and strong-coupling limits. We have tried here to shed some (faint) light on this question and plan in the future to use the simplified setting of a spherical pore to further elucidate it.

Finally, our results are obtained within the hypothesis of continuous and homogeneous media. As already explained in the introduction the profile of the dielectric constant is not uniform in the nanopore, increasing from the surface to the center~\cite{Bonthuis2012,Loche}.
This recent numerical work means that ions will be further excluded from the region close to the pore surface and forced to be located at the center. This effect can be reinterpreted in our model as a reduction  of the effective pore radius. Hence the coexistence lines in the phase diagrams in the pore radius vs. bulk concentration plane (see Figs.~\ref{PD1},\ref{PD2})  would be slightly shifted to the right (towards higher bulk concentrations). Furthermore, taking into account the ionic excluded volume in the grand-potential leads to a shift of the coexistence line towards smaller bulk concentrations because, as shown in~\cite{JCP2016}, this interaction strongly modifies the PMF in the bulk.

Furthermore to properly model the dielectric constant the solvent molecules should be modeled as self-orienting dipoles, which makes it depend in principle also on the salt concentration. This has been studied in the bulk using field-theoretic approaches~\cite{Koehl,Levy,Adar}. Developing these types of approaches to the case of a confined electrolyte would be an interesting objective for future work.

Extending our work to the case of discrete fixed charges would be an important but challenging future endeavor. The influence of such dopant charges located along the pore has been studied in simple one-dimensional nanopores by assuming a 1D Coulomb potential between charges (meaning that the electric field does not enter the membrane at all, see~\cite{Zhang2}). These discrete fixed charges can induce interesting and surprising effects phenomena, such as ion exchange phase transitions~\cite{Zhang2} or the fractional Wien effect ~\cite{Kavokine2}. Because of the complexity of the problem the case of discrete charges with a tunable dielectric constant inside the pore would, however, most likely require numerical simulation techniques.

\acknowledgments

This work was supported by the Agence Nationale pour la Recherche (project IONESCO No. ANR-18-CE09-0011-01). We are tributary to the Universities of Toulouse III - Paul Sabatier and of Montpellier and the Centre National de la Recherche Scientifique (CNRS).

\section*{Appendix}

In this appendix, we give the electrostatic potential of the studied system, composed of a charged ion (charge $q$, radius $R_i$, dielectric constant $\epsilon_i$) located at the center of a spherical nanopore of radius $R$. The inside of the pore is filled by an electrolyte of dielectric constant $\epsilon$ and Debye screening parameter $\kappa$ and the outside of the pore is composed of  a dielectric medium of dielectric constant $\epsilon_m$ devoid of electrolyte. For this system the electrostatic potential, $\Phi(r)$, is given, respectively within the ion, inside the pore, and outside the pore, by
\bea
\Phi_i (r) &=& \frac{q}{4\pi\epsilon_0\epsilon_1}\left[\frac{1}{r}-\frac{1}{R_i}+\frac{\epsilon_1}{\epsilon R_i}\frac{2\kappa R_i\sigma e^{\kappa (R-R_i)}+ \kappa R+1-\epsilon_m/\epsilon+ e^{2 \kappa (R-R_i)} ( \kappa R-1+\epsilon_m/\epsilon)}{ e^{2 \kappa (R-R_i)} (\kappa R_i+1) ( \kappa R-1+\epsilon_m/\epsilon)-(\kappa R_i-1) ( \kappa R+1-\epsilon_m/\epsilon)}\right] \label{eq:phi1}\\
\Phi (r) &=& \frac{q}{4\pi\epsilon_0\epsilon r} \frac{e^{\kappa (R-r)} \left[ e^{\kappa (R-R_i)} (\kappa R-1+\epsilon_m/\epsilon)+ \sigma (\kappa R_i-1)\right] + e^{\kappa (r-R_i)} \left[\kappa R+1-\epsilon_m/\epsilon +  \sigma (\kappa R_i+1) e^{\kappa (R-R_i)}\right]}{e^{2 \kappa (R-R_i)} (\kappa R_i+1) (\kappa R-1+\epsilon_m/\epsilon)- (\kappa R_i-1) ( \kappa R+1-\epsilon_m/\epsilon)}  \label{eq::phi2}\\
\Phi_m (r) &=&\frac{q}{4\pi\epsilon_0\epsilon r}  \frac{\sigma \left[\kappa R_i-1+ e^{2 \kappa (R-R_i)}(1+\kappa R_i) \right]+2 \kappa R e^{\kappa (R-R_i)}}{ e^{2 \kappa (R-R_i)} (\kappa R_i+1)(\kappa R-1+\epsilon_m/\epsilon)-(\kappa R_i-1) ( \kappa R+1-\epsilon_m/\epsilon)}.
\eea
For the general case of finite size ions the link between $\Phi(r)$ (Section II) and $v_0(\br,\br';\kappa_v)$ (Section III) is
\be
z^2v_0(\br,\bzero;\kappa_v) = \beta q \Phi(r).
\ee

The dimensionless electrostatic Green function (or potential) governing the interaction between an elementary point charge located at point $\br'$ and another one at point $\br$ (such that $|\br'|<|\br|$) in a neutral spherical pore can be computed exactly for point-like ions~\cite{Curtis2005} and is given (in units of $k_BT$) by
\be
v_0(\br,\br')=\sum_{l,m}V_l(r,r') Y_{lm}(\theta, \phi)Y_{lm}^*(\theta', \phi')
\label{pointlike}
\ee
where $Y_{lm}$ are the spherical harmonics and
\be
V_l(r,r')=4\pi \kappa\ell_B\frac{ \epsilon_b}{\epsilon}i'_l(\kappa r')\left[k_l(\kappa r)-\frac{\kappa R\  k'_l(\kappa R)+\frac{\epsilon_m}{\epsilon} k_l(\kappa R)}{\kappa R\  i'_l(\kappa R)+\frac{\epsilon_m}{\epsilon} i_l(\kappa R)}i_l(\kappa r)\right]
\ee
with $i_l(x)=\sqrt{\pi/(2x)}I_{l+1/2}(x)$, $k_l(x)=\sqrt{2x/\pi}K_{l+1/2}(x)$ where $I_l$ and $K_l$ are the modified Bessel functions of the first and second kinds, respectively, and $i'(x)$ and $k'(x)$ are their derivatives.

The case $l=0$ corresponds to an ion located at the center of the spherical nanopore for which the potential, $v_0(r,0)$, is isotropic and $\br'=0$.  Using  $k_0(x)=e^{-x}/x$ and $i_0(x)=\sinh(x)/x$, $v_0(r,0)$ simplifies to
\be
v_0(r,0)=\ell_B\frac{\epsilon_b}{\epsilon}\frac{e^{-\kappa r}}{r}+2\ell_B\frac{\epsilon_b}{\epsilon}D(\epsilon_m/\epsilon, \kappa R) \frac{\sinh(\kappa r)}{r}
\label{v0PL}
\ee
where
\be
D(a,x)=\frac{x+1-a}{e^{2x}(x-1+a)+x+1-a}
\ee
Hence we recover $z^2 v_0(r,0)=\beta q \lim_{R_i\to0} \Phi(r)$ for point like ions and $\sigma=0$.

From \eq{v0PL} on obtains directly for point like ions
\bea
\mu-\mu_b&=&\frac{z^2}2 \lim_{r\to0} [v_0(r,0)-v_c(r)]+\frac{z^2\kappa_b\ell_B}2\\
&=&\frac{z^2\ell_B}2\left[\frac{\epsilon_b}{\epsilon}2\kappa D(\epsilon_m/\epsilon, \kappa R)+\kappa_b-\frac{\epsilon_b}{\epsilon}\kappa+\left(\frac{\epsilon_b}{\epsilon}-1\right) \lim_{r\to0} \frac1r \right]
\eea
where we identify the three contributions for point-like ions corresponding to those of $\Delta W_{\rm p}(\kappa)$ given in \eq{W} for finite size ones. The  final term (Born self-energy) for point-like ions formally diverges when $\epsilon \neq \epsilon_b$, which shows the crucial importance of taking into account the finite ion size in this case.
For $\epsilon=\epsilon_m$ (no dielectric jump), the first term in the brackets simplifies to $\frac{\epsilon_b}{\epsilon}\kappa[1-\tanh(\kappa R)]$ (confinement solvation term) and vanishes for $R\to\infty$.

\end{document}